\documentclass[journal=jctc,manuscript=article,layout=twocolumn]{achemso}

\usepackage[T1]{fontenc}
\usepackage[latin9]{inputenc}
\usepackage[english]{babel}
\usepackage{amsmath}
\usepackage{amssymb}
\usepackage{wasysym}
\usepackage{graphicx}
\usepackage{xcolor}
\usepackage[export]{adjustbox}
\usepackage{multirow}     
\usepackage{soul}
\usepackage{subfigure}
\usepackage{tabularx}
\usepackage[para,online,flushleft]{threeparttable}
\usepackage{stackengine}
\usepackage{braket}
\setstackEOL{\\}

\newcommand{\dd}{{\rm d}}
\newcommand{\br}{{\bf r}}

\newcommand{\be}{\begin{equation}}
\newcommand{\ee}{\end{equation}}
\newcommand{\ba}{\begin{eqnarray}}
\newcommand{\ea}{\end{eqnarray}}
\newcommand{\baa}{\begin{align}}
\newcommand{\eaa}{\end{align}}

\newcommand{\lb}{\label}

\newcommand{\op}[1]{\hat {#1}}
\newcommand{\sop}[1]{\op{\op {#1}}}
\newcommand{\commutator}[2]{\left[ {#1} , {#2} \right]}
\newcommand{\trace}[1]{\mathrm{Tr}\left(#1\right)}
\newcommand{\argument}[1]{\ifthenelse{\isempty{#1}{}}{}{(#1)}}

\newcommand{\dmnot}{\op{\rho}_0}
\newcommand{\dm}{\op{\rho}}
\newcommand{\hnot}{\op{H}_0}

\newcommand{\Liouv}{\sop{\mathcal L}}

\newcommand{\identity}{\op{\mathbb I}}

\newcommand{\ICL}{Department of Materials, Imperial College London, London SW7 2AZ, United Kingdom}
\newcommand{\ICTP}{The Abdus Salam International Centre for Theoretical Physics, Condensed Matter and Statistical Physics, 34151 Trieste, Italy}
\newcommand{\CEA}{Univ.\ Grenoble Alpes, CEA, IRIG-MEM-L\_Sim, 38000 Grenoble, France}
\newcommand{\Bristol}{Centre for Computational Chemistry,
School of Chemistry, University of Bristol, Bristol BS8 1TS, United Kingdom}


\SectionNumbersOn

\title{Transition-Based Constrained DFT for the Robust and Reliable Treatment of Excitations in Supramolecular Systems}

\author{Martina Stella}     
\affiliation{\ICL}
\altaffiliation{\ICTP}
\author{Kritam Thapa}  
\affiliation{\ICL}
\author{Luigi Genovese}
\affiliation{\CEA}
\author{Laura E.\ Ratcliff}
\email{laura.ratcliff@bristol.ac.uk}
\affiliation{\ICL}
\altaffiliation{\Bristol}

\begin{document}


\begin{abstract}
Despite the variety of available computational approaches, state-of-the-art methods for calculating excitation energies such as time-dependent density functional theory (TDDFT), are computationally demanding and thus limited to moderate system sizes. Here, we introduce a new variation of constrained DFT (CDFT), wherein the constraint corresponds to a particular transition (T), or combination of transitions, between occupied and virtual orbitals, rather than a region of the simulation space as in traditional CDFT. We compare T-CDFT with TDDFT and $\Delta$SCF results for the low lying excited states (S$_{1}$ and T$_{1}$) of a set of gas phase acene molecules and OLED emitters, as well as with reference results from the literature. At the PBE level of theory, T-CDFT outperforms $\Delta$SCF for both classes of molecules, while also proving to be more robust. For the local excitations seen in the acenes, T-CDFT and TDDFT perform equally well. For the charge-transfer (CT)-like excitations seen in the OLED molecules, T-CDFT also performs well, in contrast to the severe energy underestimation seen with TDDFT. In other words, T-CDFT is equally applicable to both local excitations and CT states, providing more reliable excitation energies at a much lower computational cost than TDDFT. T-CDFT is designed for large systems and has been implemented in the linear scaling BigDFT code. It is therefore ideally suited for exploring the effects of explicit environments on excitation energies, paving the way for future simulations of excited states in complex realistic morphologies, such as those which occur in OLED materials. 
\end{abstract}

\maketitle

\section{Introduction}\label{sec:introduction}

\begin{figure}[!h]
\centering
\includegraphics[width=0.5\textwidth]{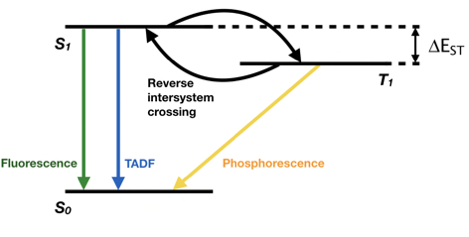}
\caption{Simplified Jablonski diagram for TADF emitters. Reverse intersystem crossing, represented in the figure by the arrow going from T$_{1}$ to S$_{1}$ (with the intersystem crossing going from S$_{1}$ to T$_{1}$), is thermally activated when the energy difference between the two excited states is smaller or comparable to $k_{b}$T.\label{fig:tadf}}
\end{figure}

Studying excited states in molecules and extended systems is one of the major ongoing challenges in physics, chemistry and materials science due to the complexity of the underlying electronic structure. Nonetheless, an accurate characterization of excitation energies is crucial for a fundamental understanding of systems of technological interest, e.g.\ solar cells~\cite{DeAngelis2014}, organic light emitting diodes~\cite{Tao2014, Wong2017} and chromophores in biological systems~\cite{Rubio2003}. One example of interest is thermally activated delayed fluorescence (TADF) emitters, which have gained a spotlight in recent years as a new type of organic light emitting diode (OLED)~\cite{Tao2014, Wong2017}. This is due to their promising maximum theoretical internal quantum efficiency (IQE) of 100\% \cite{Carlson1971, Merkel1973, Hiroki2012}. TADF relies on a reverse intersystem crossing (RISC) mechanism (triplet-to-singlet energy up conversion, illustrated by a simplified Jablonski diagram in Fig.~\ref{fig:tadf}) to achieve such high efficiency without employing expensive noble metal ions. TADF is however only possible at appreciable rates if the singlet-triplet splitting, $\Delta E_{\mathrm{ST}}$,  (defined in Fig.~\ref{fig:tadf}) is smaller than or comparable to $k_{b}$T, where $k_{b}$ is the Boltzmann constant and T the temperature. Therefore, accurate prediction of $\Delta E_{\mathrm{ST}}$ is a key but challenging element for designing more efficient TADF emitters. 
Excited states in TADF can be a mixture of charge transfer (CT) and local excitations (LE), while their nature can vary with both chemical structure and changes in molecular conformation~\cite{Olivier2017}.

Experiments for investigating excitations can be challenging because of factors such as technical setups and short lifetimes. In addition, the classification and interpretation of the nature of different excitations (i.e.\ valence state, CT, LE) is often based on empirical data. \emph{Ab initio} simulations therefore represent a valuable tool. However, as the example of TADF also highlights, there are many challenges that lie within the computational modelling of excited states, including the need to go beyond gas phase simulations and instead consider realistic morphologies~\cite{Olivier2018}.
In this context it is necessary to develop a methodology which is able to reliably capture the excited electronic structure, while accounting for both conformational and environmental effects of the full system, and still maintaining an affordable computational cost.

In this paper we present a new computational approach motivated by the desire to simulate excitations in large systems for applications such as TADF. In order to motivate our approach, we first present a overview of currently available \emph{ab initio} approaches for excited state calculations, many of which have been developed in recent years as a result of community efforts to provide accuracy (see e.g.\ Ref.~\cite{Jacquemin2009}) as well as precision benchmarks~\cite{setten2015} for molecular quantities beyond the ground state.

\subsection{Density Functional Theory-Based Methods for Simulating Excited States}

Density functional theory (DFT)~\cite{Hohenberg1964,Kohn1965} has established itself as one of the most promising approaches for studying excitations in molecules and large systems, mainly due to its notoriously favourable trade-off between accuracy and computational cost. However, it is well known that, in its standard form, DFT falls short when describing excited states because of the ground-state nature of its formulation. For this reason a range of different DFT-based methodologies have been developed in order to better account for excited electronic states.

$\Delta$SCF (self-consistent field)~\cite{Jones1989} is the simplest DFT-based approach for computing excitation energies. For a given excited state of interest the energy splitting is defined as $\Delta E_\mathrm{SCF}^{n} = E_{\mathrm{n}} - E_{0} $,
where $E_{0}$ is the ground state energy and $E_{\mathrm{n}}$ is the energy of an ``excited'' state, labelled by $n$, which is obtained by manually controlling the occupation of the Kohn-Sham (KS) states as the system reaches self-consistency. The $\Delta$SCF approach has been used with wide success because of its simplicity and low computational cost. It has, for a long time, been justified in cases where the excited state corresponds to the lowest state of a given symmetry~\cite{Lundqvist1976}, while its applicability has been also extended, such that it gets a formal justification in the general case~\cite{Gorling1999}.

Linear-response (LR) time-dependent DFT (TDDFT)~\cite{Gross1984,Casida1995} is the most commonly used method for investigating excitations in molecular systems, as it often provides good agreement with experiment. Despite being well established and more affordable than sophisticated post-Hartree-Fock methods such as CCSD(T) and CASSCF~\cite{Friesner2005}, LR-TDDFT nonetheless has limitations which prevent it from being feasibly employed in modelling realistic morphologies, such as those in which TADF emitters are employed. First, its computational cost is still too onerous for modelling systems larger than a few hundred atoms~\cite{Liu2015}, although various approaches have been developed for treating large systems, e.g.\ linear scaling TDDFT~\cite{Zuehlsdorff2015} and GPU-accelerated approaches~\cite{Isborn2012}, as well as subspace-based approaches~\cite{Neugebauer2007}. Second, it notoriously fails when describing CT states with routinely used semi-local functionals~\cite{Dreuw2004,Kummel2017}. The latter issue in particular has been extensively studied and a number of solutions are nowadays available, of which the most successful is the use of range-separated hybrid functionals~\cite{Savin}. However, such functionals are still not widely available and can make calculations more expensive, while good performance often necessitates the tuning of the functional parameters for the system in question~\cite{Kummel2017}.

Another DFT-based method for studying excited states is constrained DFT (CDFT)~\cite{Dederichs1984,Wu2005}. In CDFT a constraint is added to the density, following which the energy is found by minimising the density with this additional condition. In its most common form a specific electronic charge is constrained to a region of simulation space. If opportunely guessed, such a constraint can correspond to a specific excited state where the electronic density is well localized within the region, and takes into account, \emph{by design}, the self-consistent response of the system to the imposition of such a constraint. For this reason, CDFT has performed very successfully for molecular systems with an obvious spatial separation between donor and acceptor regions~\cite{Wu2005,Kowalczyk2010,Segal2007,Difley2008,Sena2011,Souza2013,Difley2011,Roychoudhury2016}. Such a simple approach naturally overcomes some of the well-known limitations of DFT, e.g.\ the self-interaction problem and the resulting delocalization errors, making CDFT particularly appropriate for treating CT states and an asset for modelling exciton formation. CDFT is conceptually intuitive and follows the same scaling as DFT (generally scaling with the cube of system size). However, it is most appropriate where some information is known about the excitation in question (where and how much charge to impose). 
For a comprehensive review of CDFT, we refer the reader to Ref~\cite{Kaduk2012}.

The simplicity of its framework has also made CDFT attractive for the development of new variations. Recently, Ramos and Pavanello~\cite{Ramos2016, Ramos2018} proposed two versions of CDFT. In a first implementation~\cite{Ramos2016} they combine CDFT with a frozen density embedding approach. The method, termed constrained subsystem DFT (CSDFT), is mainly applied to describe hole transfer reactions. In a later paper~\cite{Ramos2018}, 
they present a CDFT method tailored to compute low-lying electronic excitations (XCDFT) of molecular systems, which resolves the space of virtual states by projection. The results show an accuracy only slightly worse than LR-TDDFT. A more recent paper by Roychoudhury \emph{et al.}~\cite{ORegan2020} proposes a generalization of CDFT for charge-compensating electronic excitations in molecules (XDFT). The obtained results are again comparable to TDDFT.

Beyond the above methods, there are also other DFT-based approaches to excited state calculations, which are either generally applicable, such as orthogonality constrained DFT~\cite{Evangelista2013}, or designed for CT states, such as constricted variational density functional theory~\cite{Senn2016}.
In short, there are many approaches which can compute low lying excitations in molecular systems, with each displaying limitations either in the ability to describe particular classes of excited states (LE \emph{vs.}\ CT), or in the maximum accessible system size. 

Furthermore, beyond the ability to treat many atoms, the complexity of large systems often also necessitates the ability to map to local degrees of freedom (DoF)~\cite{Mohr2017b,Dawson2020}. To this end, it would be highly advantageous to have an excited state method which can be related to a local description of a large system, e.g.\ to excite a single molecule within a cluster of molecules. Given all these factors, there is no clear consensus on the best approach to use, particularly for applications such as TADF where the nature of the excitation can be a combination of LE and CT, and the effect of an extended environment can be crucial for accurately describing the excitation. Nonetheless, the recent variants of CDFT demonstrate its potential for providing accurate results with a lower computational cost than TDDFT. 

In this paper we present an alternative variation of CDFT. In our approach, which is implemented in the wavelet-based BigDFT code~\cite{Ratcliff2020b}, the constraint is defined as a particular \emph{transition}, or combination of transitions, between given occupied and virtual states, rather than a region of the simulation space. The approach is therefore termed transition-based CDFT (T-CDFT). The transition constraint takes inspiration from an optical excitation term, rigorously defined from LR  equations and parameterized for instance from LR-TDDFT. In this context, it can be considered as a further step beyond LR calculations, where SCF effects are added \emph{on top} of the optical excitation. We consider both `pure' transitions between one occupied molecular orbital (the highest occupied molecular orbital) and a single orbital in the unoccupied sector -- which require no additional simulation input -- as well as `mixed' excitations involving more than one occupied $\rightarrow$ virtual transition, where we use TDDFT to define the transition breakdown.  We benchmark our approach on low lying singlet and triplet excitations of a set of molecules in gas phase, including both acenes and OLED emitters, putting the results in comparison with $\Delta$SCF and TDDFT calculations.  While the present work focuses on gas phase simulations, we also describe a path towards future large scale simulations, using the ability of BigDFT to treat systems containing thousands of atoms~\cite{Mohr2014,Mohr2015}.

The outline is as follows. In Section~\ref{sec:methods} we first present the underlying formalism of T-CDFT, before describing the implementation in BigDFT. We finish Section~\ref{sec:methods} by defining two indicators which will be used for analyzing excitations, and specifying the computational details. In Section~\ref{sec:results} we present the results, first discussing the nature of the excitations, including both LE/CT character and the effect of treating excitations as pure or mixed. We finish with a detailed comparison between the obtained excitation energies for the different methods, for both LE and CT excitations. Finally, in Section~\ref{sec:conclusions}, we conclude. 

\section{Methods}\label{sec:methods}

\subsection{Excitations in the Linear-Response Formalism}

When a system is submitted to an excitation, its density matrix, and therefore its
observables,  are modified by the effect of a (potentially frequency-dependent) perturbing operator $\op\Phi(\omega)$, with $\omega$ being the frequency.
Stated otherwise, we can identify a response density operator $\dm'(\omega)$ which 
represents the deviation of the density matrix from the ground-state equivalent, indicated by $\dmnot$.
Such a response density satisfies an equation of motion written in the form of a quantum Liouville (super) operator, $\Liouv$, (see for example Ref.~\cite{Rocca2008}):
\be\lb{LiouvillianRhopomegaDef1}
\left(\omega - \Liouv\right) \dm'_\Phi(\omega) =  \commutator{\op\Phi(\omega)}{\dmnot} \;.
\ee
Its action on a generic operator $\op O$ reads
\be\lb{LiouZeroDef1}
\Liouv \op O \equiv \commutator{\hnot}{\op O} +\commutator{\op V'[\op O]}{\dmnot}  \;,
\ee
where $\hnot$ is the ground-state Hamiltonian and $\op V'[\op O] \equiv \int \dd \br \dd \br'\frac{\delta \op V[\dmnot]}{\delta \rho(\br,\br')} O(\br,\br')$ encodes the response of the $\dm$-dependent potential to a modification of the density operator.
The ``excitation modes''  of the molecule are defined through the \emph{excitation operators} $\op E_a$, satisfying
\be\lb{ExcitationOperatorsDef1}
\Liouv \op E_a = \Omega_a \op E_a \;,
\ee
with $\Omega_a$ being the excitation energies.
The operator orthonormalization condition 
\be\lb{orthoExcitatioOpDef1}
\trace{\op{\tilde E}_a\op E_b} = \delta_{ab} \;,
\ee
where $\op{\tilde E}_a$ is the excitation operator associated to the left Liouvillian eigenproblem~\cite{D'Aless2019}, 
guarantees that the excitation operators can be seen as a basis for representing the perturbation of the system.

Under a linear-response condition, it is possible to show \cite{D'Aless2019} that the excitation operators satisfy the so-called \emph{transverse} condition, 
which states that
\be\lb{RhopTransverseDef1}
\left(\op E_a\right)_\perp \equiv
\dmnot \op E_a \op Q_0 + \op Q_0  \op E_a \dmnot  = \op E_a\;,
\ee
where $\op Q_0=\identity-\dmnot$ is the projector to the empty subspace of 
the ground-state Hamiltonian $\hnot$.
According to this condition, excitation operators can be parametrized as
\begin{align}\lb{ExcitationOperatorsDef2}
\op E_a &= \sum_p \left( \ket{\phi^a_p}\bra{\psi_p} + \ket{\psi_p} \bra{\chi^a_p}\right)\;.
\end{align}
We here indicate as $
\psi_p$ the occupied orbitals, where $
\phi_p^a$ and $
\chi_p^a$ are associated to vectors belonging to the span of unoccupied states (and are therefore orthogonal to the set of all occupied orbitals).

Each excitation mode of the system, with associated energy $\Omega_a$, is thus described by a set of states $\{\phi^a_p,\chi^a_p\}$, each defined in the unoccupied subspace. It is possible to show~\cite{D'Aless2019} that these objects represent, respectively, the state into which $\ket{\psi_p}$ is excited -- or from which it decays -- when the system is subject to the \emph{monochromatic} perturbation $\op \Phi_a \equiv \commutator{\op E_a}{\dmnot}$,
which would only resonate with the excitation having an energy $\Omega_a$ (see~\cite{D'Aless2019}). The spectrum is symmetric with respect to the inversion of the eigenvalues $\Omega_a \rightarrow -\Omega_a$ and, given a specific excitation $\{\phi^a_p,\chi^a_p\}$,
the associated solution with opposite energy is described by the transposed pair $\{\chi^a_p,\phi^a_p\}$.

\subsection{Transition-Based Constrained DFT}\label{sec:cdft}

Excited states, as calculated for example with LR-TDDFT, may be characterized by the orbitals involved in a given transition. Each of the occupied orbitals labelled by $\psi_p$ is then associated with a particular transition of this state in the unoccupied sector.

Following these guidelines, we define the (Hermitian) transition operator, $\op T_a$:
\begin{multline}
\op T_a\equiv \frac{1}{\sqrt 2} \left(\op E_a + \op E_a^t\right) = \\ = \sum_p \frac{1}{\sqrt 2}\left( \ket{w^a_p}\bra{\psi_p} + \ket{\psi_p} \bra{w^a_p}\right)\;,
\end{multline}
which is associated with the linear combination of an excitation and a de-excitation
with the same energy. Normalization of the excitation modes implies that $
\sum_p \braket{w^a_p | w_p^a} =1$.
A representation of the states $\ket{w_a^p}\equiv \ket{\phi^a_p} + \ket{\chi^a_p}$ can be provided by introducing
an explicit basis $\{\ket{s}\}$
in the subspace of empty states so that both $\ket{\phi^a_p}$ and $\bra{\chi^a_p}$ can be represented as
\be
\ket{w_p^a}=\sum_s W^a_{p s}\ket{s} \;.
\ee
The representation of the above equation in the basis set of the eigenvalues of the ground-state Hamiltonian gives rise to the TDDFT Casida's equations.
For semi-local DFT functionals, the normalized coefficients $W^a_{p s}$ can be directly extracted from the Casida's coupling matrix eigenproblem, thereby providing an explicit representation of $\ket{w_p^a}$. 

We define our excitation energies \emph{via} the following:
\be
E_{\mathrm{T-CDFT}}^{(a)} \equiv  E[\rho]\biggr|_{\trace{\dm \op T_a} = 1} - E[\rho_0]\;,
\ee
where we denote by $E[\rho]$ the SCF energy obtained from the density $\rho$. In other terms, we minimize the energy by imposing self-consistency under
the transition constraint imposed by the operator $\op T_a$.
More specifically, the density matrix operator is constrained such as to include the transitions $\psi_p \rightarrow w_p^a$
in its definition. The energy of the system is then minimized along the set of solutions implementing such a constraint. 

\subsubsection{Transition Breakdown}

We now describe the procedure for imposing the constraint onto the density.
A particular excitation (labelled by $a$) is characterized by the set of occupied states $\ket{\psi_p}$ which are excited into 
the corresponding unoccupied orbitals $\ket{s}$ with a weight provided by the coefficients $W^a_{p s}$. 
This may involve only one occupied $p$ orbital, e.g.\ the highest occupied molecular orbital (HOMO), or it may involve a mixture of several orbitals.  We refer to the former as a `pure' transition, while the latter is considered a `mixed' transition. For a given orbital $p$ the level of purity of the transition $a$ can be quantified by means of a transition purity indicator, $\mathcal{P}_p^a$, defined as
\be\label{eq:tpi}
\mathcal P_p^a \equiv \braket{w_p^a | w_p^a}\,;
\ee
which would lead to a value of one if only the orbital $\psi_p$ participates in the transition. Clearly $\sum_p \mathcal P_p^a =1 $ for any excitation $a$.

For the generic case of a mixed transition, it is then enough to split the T-CDFT approach into multiple constraints:
by decomposing the transition $\op T_a$ into partial terms:
\begin{equation}
 \op T_a = \sum_p \op T_p^{(a)}\;,
\end{equation}
where $\op T_p^{(a)}$ is a pure transition defined from the sole ket $\ket{w_p^a}$, we can define the energy of the mixed transition by the SCF energy obtained from the 
density operator
\begin{equation}
  \op \rho^a = \sum_p \mathcal P_p^a \op \rho_p^a\;,
\end{equation}
where $\op \rho_p^a$ is the SCF density obtained from the pure T-CDFT calculation with the constraint $\trace{\dm \op T_p^{(a)}} = 1$.

\subsubsection{Extracting Singlet and Triplet Excitations}

With this representation, the orbital sector of the configuration interaction space is, by construction, identical between up and down spins. Therefore, spin-contamination is forbidden, and we work in the subspace of $S_z=0$ excitations. Singlet (+) and triplet ($-$) transitions can be easily identified, taking the solutions with $W^a_{ps\uparrow} = \pm W^a_{ps\downarrow}$, respectively.

In the KS DFT formalism, $\dmnot$ is found by SCF optimization of the following Lagrangian:
\begin{equation}
    E_{BS} = \trace{\hat H_{\mathrm{KS}}[\dm] \dm} \;,
\end{equation}
where the KS Hamiltonian $H_{\mathrm{KS}}$ is functionally dependent on the density matrix.
To impose the constraint we may add to the above Lagrangian the following term
\begin{equation}
    C_a = \sum_{p,s}  V_a^{ps} \left(\bra{s} \dm \ket{\psi_p} - W^a_{ps} \right)\;.
\end{equation}
The set of Lagrange multipliers $V_c^{s,s'}$ is there to enforce the condition
\begin{multline}
    \trace{\dm \op T_a} = \sum_{p,s} \left( W^a_{ps} \bra{s} \dm \ket{\psi_p} \right)= \\ =\sum_p \bra{w_p^a} \dm \ket{\psi_p} =1
\end{multline}
which would add to the KS Hamiltonian a density-independent term:
\begin{equation}
   \hat{H}_c^a = \sum_{p,s} V_a^{ps} 
   \left( \ket{\psi_p} \bra{s} + \ket{s} \bra{\psi_p}\right)\;,
\end{equation}
where the sum should only be performed on the set of states $p,s$ such that $W^a_{ps} \neq 0$. Once the appropriate values for the Lagrange multipliers are found,
the energy of the excited system can be calculated with the usual KS expression, after removing the constraining term from $H_{\mathrm{KS}}$. Singlet and triplet transitions can then be associated with
constraint operators which are spin-averaged (+) and spin-opposite ($-$), respectively: (${\hat{H}_c^{a\uparrow}}= \pm {\hat{H}_c^{a\downarrow}}$). 

Such a formalism is particularly practical because for each pure transition $\op T_p^{(a)}$, one can choose an arbitrarily large Lagrange multiplier, as the representability condition of the density matrix will always guarantee $\bra{w_p^a} \dm \ket{\psi_p} \leq 1$. A value of a magnitude of $20$ atomic units
for $V_c$ is largely sufficient for imposing the constraint, as discussed in Section~\ref{sec:computational_details}. Therefore, although the computational cost increases with the number of components in a given mixed excitation, the overall cost is comparable to traditional CDFT, where several calculations may be required to identify the Lagrange multiplier which satisfies the constraint.

\subsection{Indicators for Analyzing Excitations}\label{sec:indicators}

When analyzing excited states, it is useful to employ quantitative indicators which allow the comparison of various features of a given excitation.  This includes the orbitals involved in a given transition, and the spatial character.  To this end, we define the following two simple indicators.

\subsubsection{Transition Purity}

Both the transition purity indicator, $\mathcal{P}$ (Eq.~\ref{eq:tpi}), and the T-CDFT formalism may be applied to any transition. 
In this work, a $\mathcal{P}=1$ refers to a pure HOMO-virtual excitation, while a number significantly below 1 indicates that excitations involving deeper occupied orbitals have a non-negligible contribution to the overall description of the excitation. 

\subsubsection{Spatial Overlap}

The accuracy of computed excitation energies strongly depends on whether the chosen method is able to correctly capture the character of the excitation.  As discussed, this is particularly true for TDDFT, where energies of low-lying CT states calculated with some functionals may be severely underestimated, in some cases by several eV, while other functionals agree well with experiment~\cite{Jacquemin2008}. 
This has motivated the development of diagnostic tools for predicting the accuracy of TDDFT in a given case by classifying the nature of the excitation.  It is not straightforward to develop a unique and effective descriptor, so that several examples of such developments can be found in the literature~\cite{Adamo2011,Bedard2010,Nitta2012,Peach2008,Plesser2012}. These generally include geometric descriptors based on the molecular orbitals and electron densities. 

One such descriptor is the $\Lambda$-index~\cite{Plesser2012,Guido2013}, which is based on the overlap of molecular orbital moduli, and is defined as
\begin{equation}
\Lambda^a = \sum_{ps}{W_{ps}^a}^{2} \int{\left|\psi_s(\mathbf{r})\right|\left|\psi_p(\mathbf{r})\right|}\,\,\mathrm{d}\mathbf{r}. \label{eq:CT_1}
\end{equation}
It has been suggested that a small orbital overlap, defined by a $\Lambda$ value below 0.3-0.4, depending on the functional, corresponds to a CT transition which  is not correctly described by TDDFT~\cite{Guido2013}.

While it is possible to define indicators such as $\Lambda$ which depend on the output of the excited state calculation, in this work we instead employ a simplified descriptor which is based on the particular transition that is used as the constraint in T-CDFT, i.e.\ the simulation input, rather than the output. For the case of pure HOMO-LUMO (lowest unoccupied molecular orbital) transitions, this results in a simplified version of the $\Lambda$-index. This descriptor, which we denote $\Lambda_{\mathrm{T}}$, is based on the square root of the product of the overlap of solely the HOMO and LUMO wavefunctions and thus describes their \emph{spatial overlap}:
\begin{equation}
\Lambda_{\mathrm{T}} = \int\left|\psi_\mathrm{HOMO}(\mathbf{r})\right|\left|\psi_\mathrm{LUMO}(\mathbf{r})\right|\,\,\mathrm{d}\mathbf{r}.
\label{eq:CT_3}\end{equation}
At the two extremes, a value of zero indicates no spatial overlap between the HOMO and LUMO, and hence a CT excitation, while a value of one represents full spatial overlap, corresponding to a LE state. We note that $\Lambda_\mathrm{T}$ does not distinguish between singlet or triplet excitations, nor does it take into account additional contributions in the case of mixed excitations. For work requiring an in-depth analysis of a particular excitation, a modified version based on the output of T-CDFT, or other indicators such as those referenced above would therefore be more appropriate.

\subsection{Implementation in BigDFT}\label{sec:implementation}

We have implemented T-CDFT in the BigDFT code~\cite{Ratcliff2020b}, which uses a Daubechies wavelet basis~\cite{Daubechies1992}.  By taking advantage of the orthogonality, compact support and smoothness of wavelets, and in conjunction with accurate analytic pseudopotentials (PSPs), BigDFT is able to yield a high, systematically controllable precision.  It has both a standard cubic scaling approach with respect to the number of atoms~\cite{Genovese2008}, and a linear scaling (LS) algorithm which can treat thousands of atoms~\cite{Mohr2014,Mohr2015}. The T-CDFT implementation builds on the existing CDFT implementation in LS-BigDFT~\cite{Ratcliff2015a,Ratcliff2015b}, wherein a support function (or wavefunction) basis is constructed for the ground state, then used as a fixed basis for the (T-)CDFT calculation.
In the following we therefore first summarize the support function (SF)-based approach employed in LS-BigDFT. We then describe the generation of both SF and extended KS wavefunction basis sets, where the latter is used to verify the suitability of the SF basis for excited states.

\subsubsection{Linear Scaling BigDFT}

In LS-BigDFT the extended KS orbitals are expressed in terms of a set of localized SFs, $\phi_{\alpha}$,  \emph{via} the coefficients $c^{\alpha}_{i}$,
\be\label{eq:supp_func}
|\psi_i\rangle = \sum_{\alpha}{c^{\alpha}_{i}|\phi_{\alpha}\rangle}.
\ee
The density matrix, $\dm$, is then defined in terms of the SFs and the density kernel, K$^{\alpha\beta}$~\cite{Hernandez1995},
\be\label{eq:kern}
\dm = \sum_{\alpha\beta}\ket{\phi_{\alpha}} K^{\alpha\beta}\bra{\phi_{\beta}}\;.
\ee
By taking advantage of the well known nearsightedness principle~\cite{Hernandez1995,Kohn1996},
it is possible to impose strict localization on both the SFs and density kernel. In BigDFT, the SFs are represented in the underlying wavelet basis set and optimized \emph{in situ} during the self-consistency cycle. This results in a set of localized SFs which have adapted to their local chemical environment, giving a minimal basis which, by construction, can represent the occupied KS orbitals. 
Since the SFs are truncated within a user-defined localization radius, $R_{\mathrm{loc}}$, systematic convergence is possible by increasing the localization radius.

The density kernel is then also optimized, either by means of the Fermi Operator Expansion (FOE)~\cite{Goedecker1994,Goedecker1995} approach, which works directly with the density kernel, or with direct minimization or diagonalization approaches, which are used to obtain the coefficients $c^{\alpha}_{i}$, from which the kernel can then be constructed.  FOE used in conjunction with sparse matrix algebra, as implemented in the CheSS library~\cite{Mohr2017b}, results in LS computational cost.

The localized SFs of LS-BigDFT can also be used as a means for further reducing the complexity of calculations of large systems. This is achieved \emph{via} a fragment-based analysis, in which the system is divided into subsystems, and can be used both to reduce the computational cost by exploiting similarity between fragments~\cite{Ratcliff2015a,Ratcliff2019}, and to identify independent fragments and analyze interactions between them~\cite{Mohr2017a,Dawson2020}. SF-based T-CDFT is fully compatible with these fragment-based approaches; future work will aim at exploiting this to study excitations in environments.

\subsubsection{Support Function Basis}\label{sec:sf_approach}

The use of a transition-based constraint relies on the ability to accurately represent both the occupied orbitals and the virtual orbital(s) involved in the constraint.  The SFs are optimized to describe the occupied states, however, as in the ONETEP code~\cite{Prentice2020}, which uses a similar approach to optimize the basis of non-orthogonal generalized Wannier functions (NGWFs), the unoccupied states can be poorly represented~\cite{Skylaris2005,Ratcliff2011}. In ONETEP this problem is overcome by using a projection operator to optimize a second set of NGWFs to represent the virtual states, which are then combined with the ground state NGWFs~\cite{Ratcliff2011}.

In LS-BigDFT we instead retain a single set of SFs, exploiting the direct minimization approach to optimize select virtual states alongside the occupied states~\cite{Mohr2014}.  When only a few virtual states are required, this may be done in a single calculation.  However, when a larger number of virtual states are required, it is more stable to employ a two-step approach:
\begin{enumerate}
    \item Ground state calculation for occupied states only, to obtain the density, and an initial guess for both the SFs and kernel.
    \item Non-self-consistent calculation in which the SFs and kernel are further optimized to represent a number of virtual states.
\end{enumerate}
While step 1 may employ any approach to kernel optimization, step 2 requires the use of the direct minimization approach. Since the virtual states are more delocalized than the occupied states, particularly in the case unbound of states, it is typically necessary to use larger localization radii, while depending on the nature of the virtual states, it may also be necessary to increase the number of SFs.  

\subsubsection{Wavefunction Basis}\label{sec:wfn_approach}

In order to validate T-CDFT with a SF basis, we have also employed a wavefunction (WFN)-based approach, wherein the basis set is generated \emph{via} a ground state cubic scaling calculation with a (large) number of virtual states. These wavefunctions are then used for T-CDFT, treating them as a fixed SF basis with effectively infinite localization radii.
By increasing the number of virtual wavefunctions, it is possible to systematically approach the complete basis set limit, assuming that the set of excited states $w_p^a$ is localized (which is always true for excitations below a given threshold, see~\cite{D'Aless2019}). Such an approach is therefore possible for comparing the choice of the excitation operator in different computational setups.

\subsection{Computational Details}\label{sec:computational_details}

Vertical singlet and triplet excitations were calculated using T-CDFT, $\Delta$SCF and TDDFT. T-CDFT calculations used PBE~\cite{Perdew1996} only, as hybrid functionals are not available in LS-BigDFT, while $\Delta$SCF and TDDFT calculations used both PBE and PBE0~\cite{Adamo1999}. Since T-CDFT is targeted at large systems, where hybrid functionals are often prohibitively expensive, this reflects the computational setup which would be used at scale. All ground state calculations and singlet T-CDFT calculations are spin restricted, while the remainder of the excited state calculations are unrestricted.

$\Delta$SCF energies and the spatial overlap parameter were calculated using cubic-scaling BigDFT, where excited state calculations used the ground state wavefunctions as an initial guess to avoid convergence onto local minima, as described in the Supporting Information. The $\Delta$SCF procedure notoriously models the non-Aufbau electronic singlet state which is not a spin eigenfunction~\cite{Kowalczyk2011}. To obtain the energy of the singlet excited state, we therefore applied the common spin purification formula to the uncorrected mixed state energy $E_{\mathrm{S}_1}^{\mathrm{purified}}$,
\begin{equation}\label{eq:purification}
E_{\mathrm{S}_1}^{\mathrm{purified}} = 2E_{\mathrm{S}_1}^{\mathrm{unpurified}} - E_{\mathrm{T}_1}.
\end{equation}
All reported S$_1$ $\Delta$SCF energies are $E_{\mathrm{S}_1}^{\mathrm{purified}}$.

TDDFT calculations employed NWChem~\cite{Valiev2010} using the Tamm-Dancoff approximation (TDA)~\cite{Tozer2013}, with a cc-pVTZ basis set~\cite{Dunning1989a}. LR-TDDFT calculations were also performed using BigDFT, using the full Casida formalism~\cite{Casida1995}, in order to determine the transition breakdown and purity. As only LDA~\cite{Ceperley1980} is available for TDDFT in BigDFT, these calculations, including the transition breakdowns, were performed using LDA with a WFN basis generated using PBE; the difference compared to using a basis generated with LDA was found to be negligible.  T-CDFT calculations employed a SF basis with 4/9/9 SFs for H/C/N with $R_{\mathrm{loc}}=4.23$~\AA; the calculated values were found to be within 0.05~eV of the converged WFN-based results, demonstrating that the SF basis is complete enough to allow accurate fixed-basis T-CDFT calculations, providing the virtual states of interest are well represented.  A Lagrange multipler of -20 was used for all T-CDFT calculations, as this was found to be large enough to converge $\trace{\textbf{KW}}$ -- an example convergence plot is shown in the Supporting Information, alongside further computational details.

\section{Results}\label{sec:results}

\begin{figure}[ht]
\centering
\subfigure[naphthalene]{\includegraphics[height=0.076\textwidth]{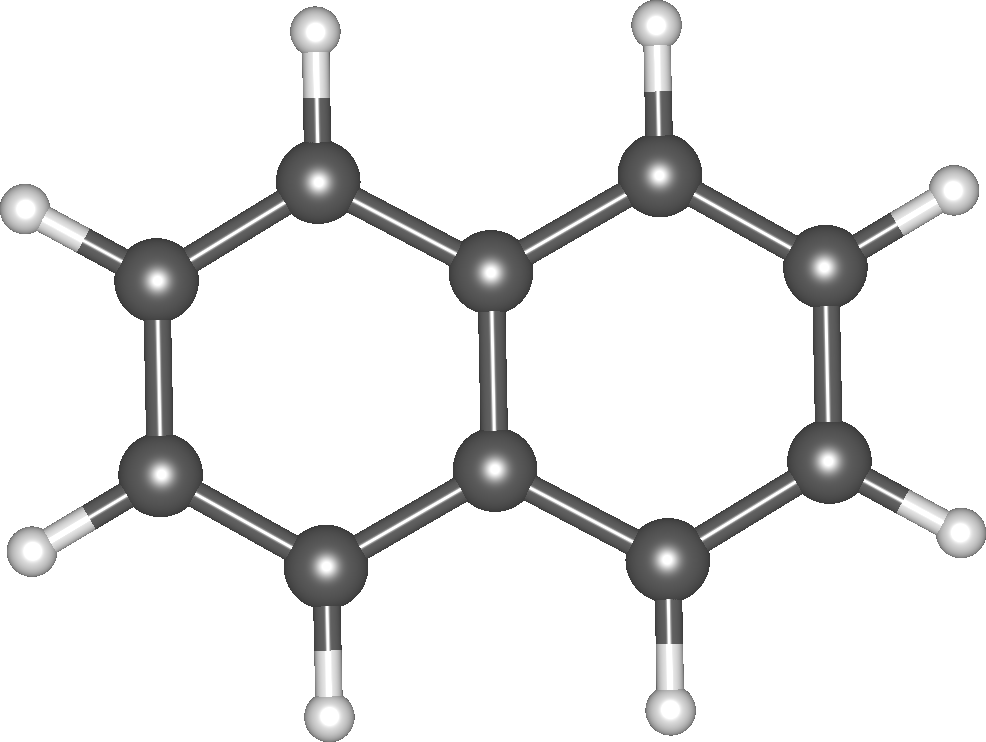}\label{fig:naphthalene}}
\hspace{2pt}
\subfigure[anthracene]{\includegraphics[height=0.076\textwidth]{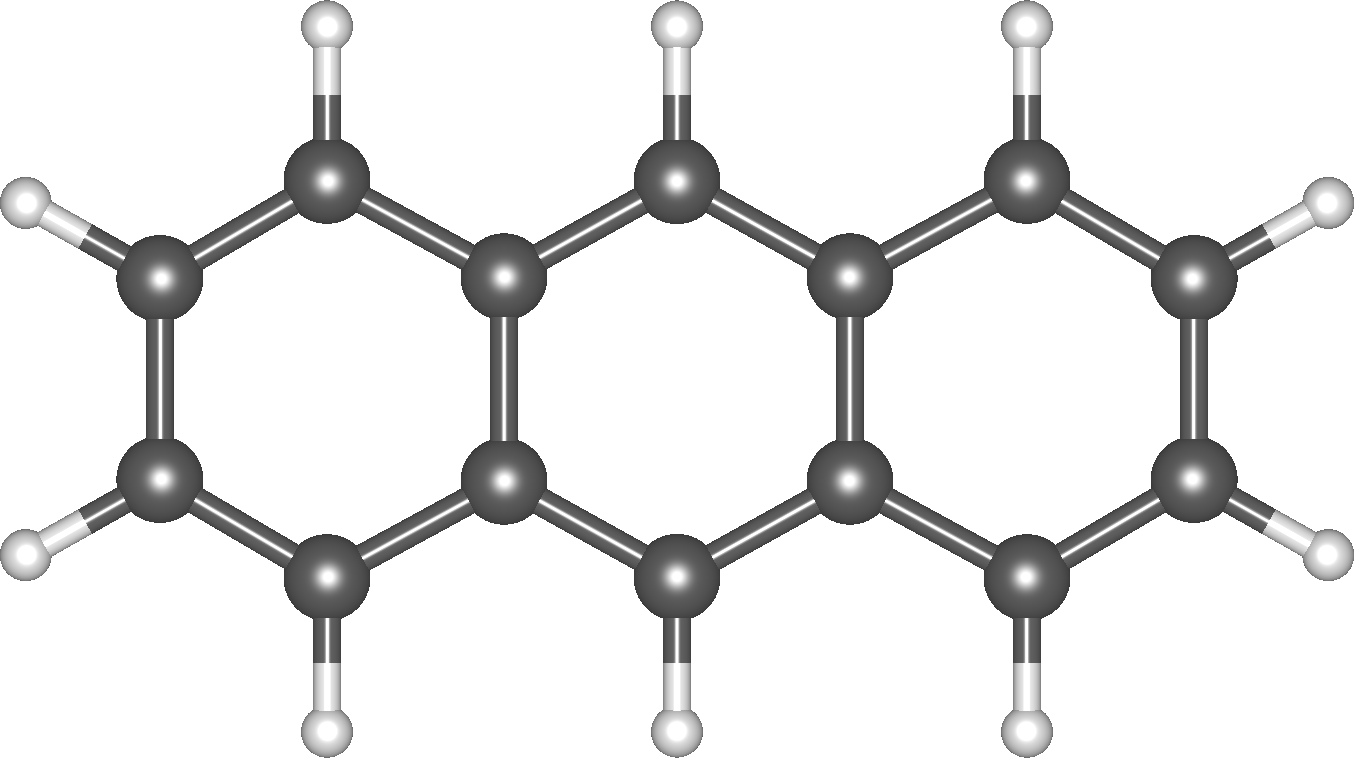}\label{fig:anthracene}}
\hspace{2pt}
\subfigure[tetracene]{\includegraphics[height=0.076\textwidth]{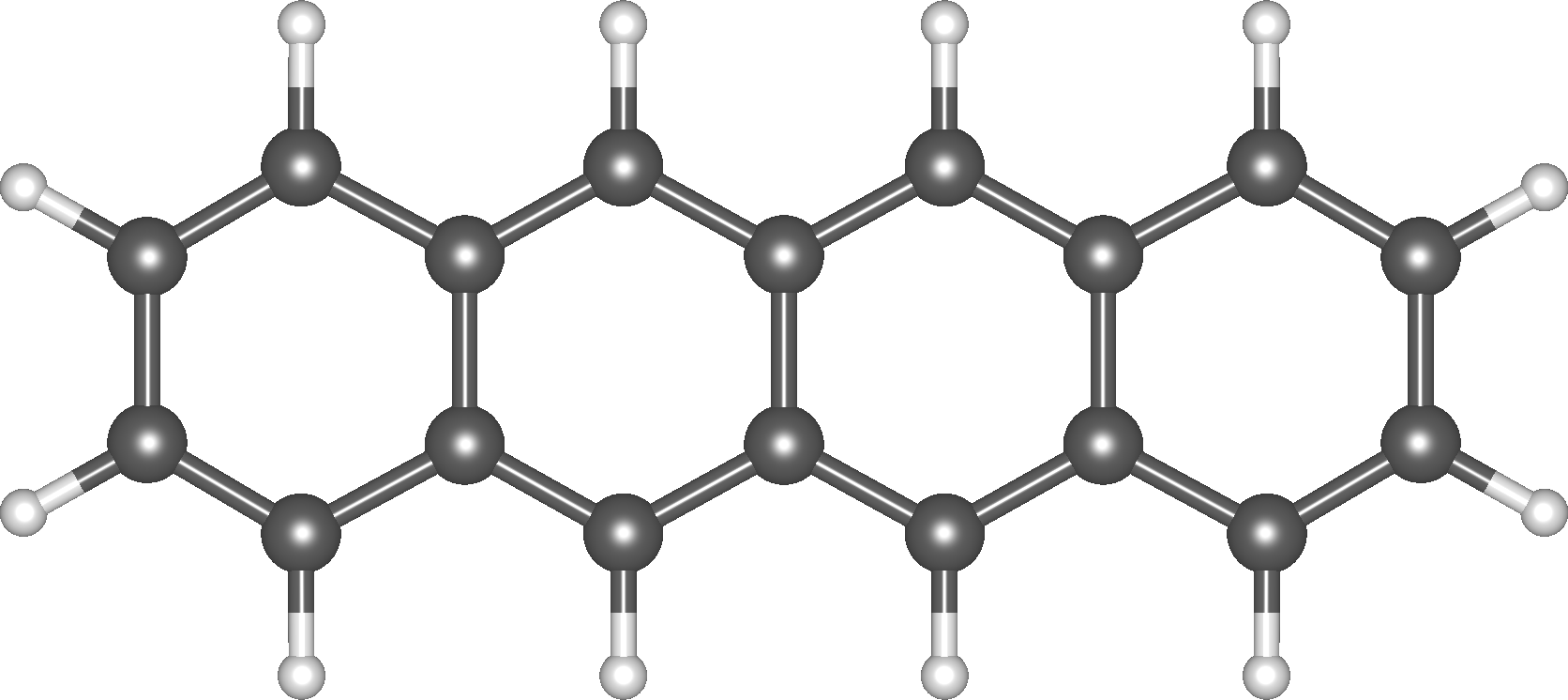}\label{fig:tetracene}}
        
\subfigure[pentacene]{\includegraphics[height=0.076\textwidth]{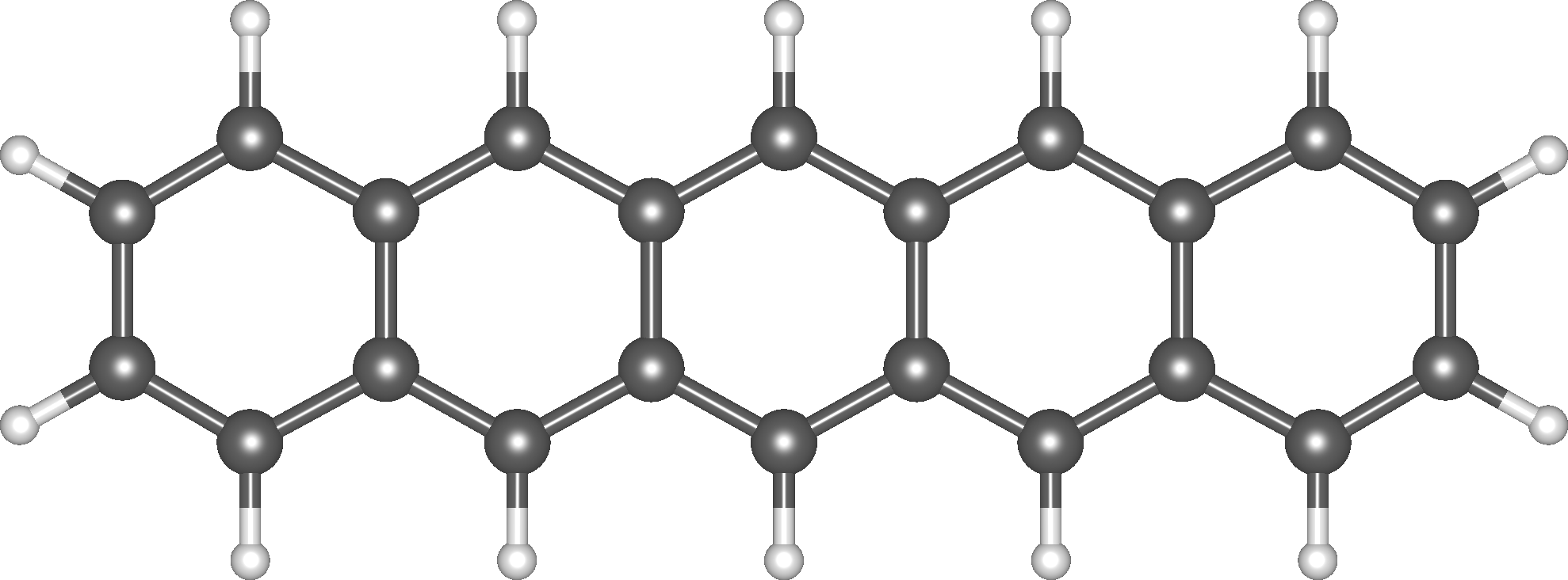}\label{fig:pentacene}}
\hspace{2pt}
\subfigure[hexacene]{\includegraphics[height=0.076\textwidth]{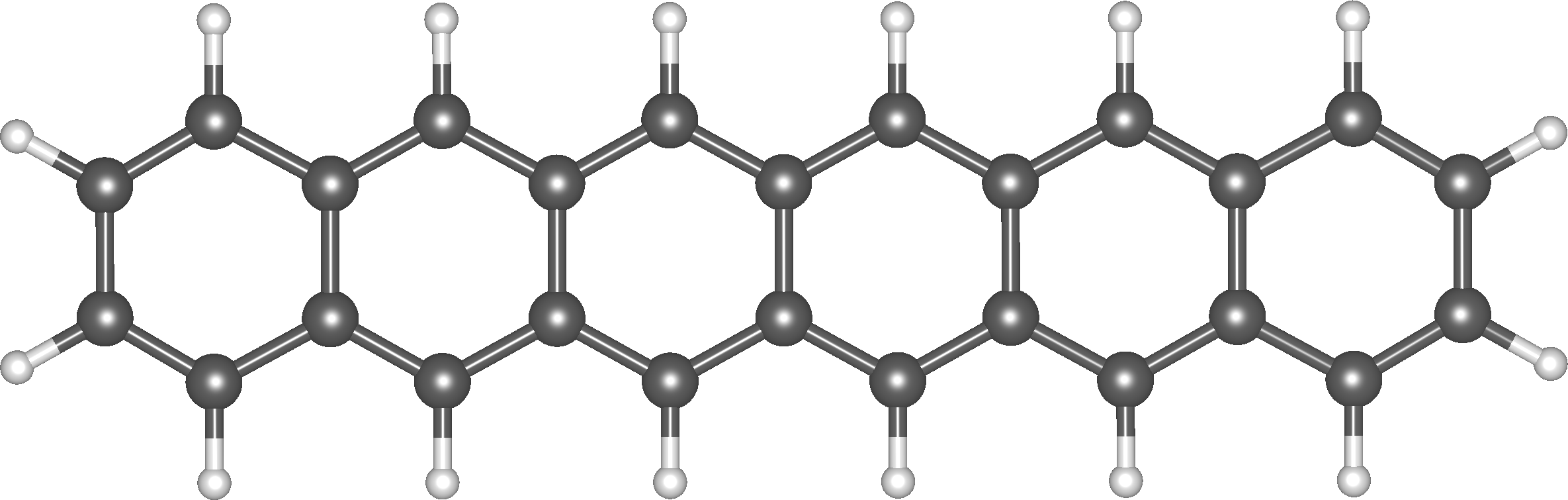}\label{fig:hexacene}}
        
\subfigure[NPh$_3$]{\includegraphics[width=0.13\textwidth]{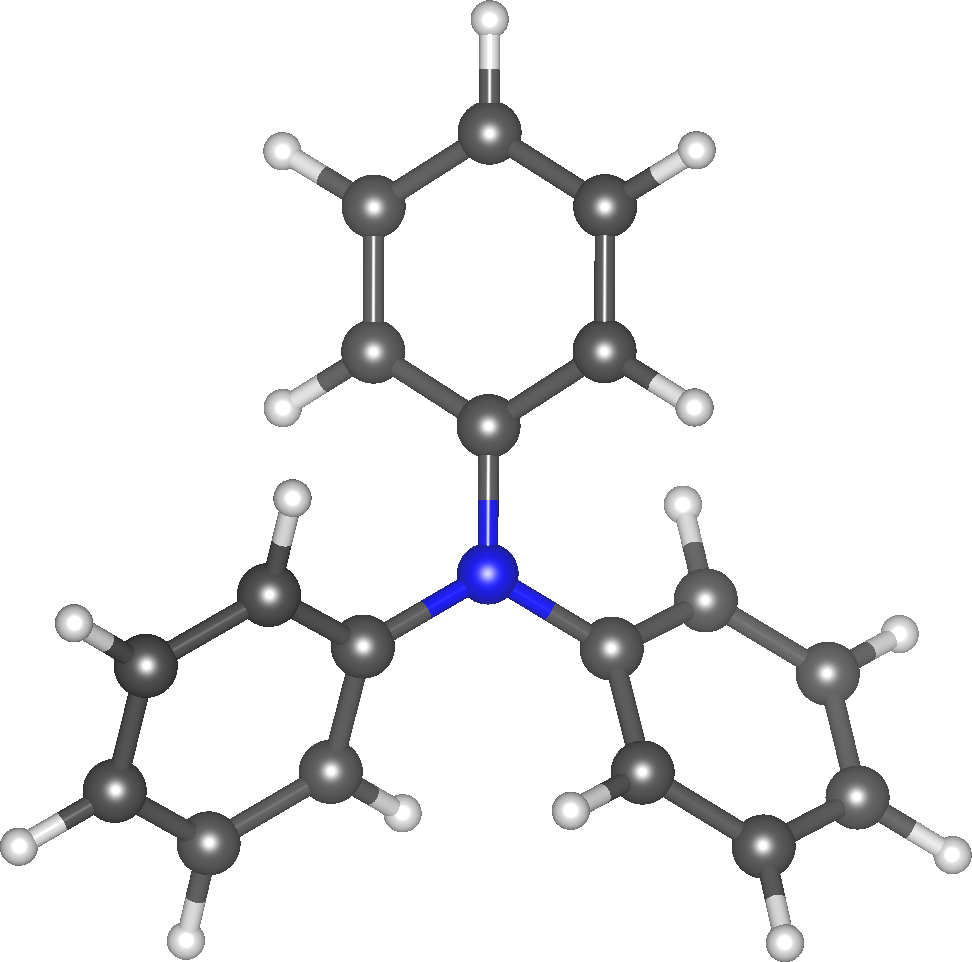}\label{fig:nph3}}
\hspace{2pt}
\subfigure[2CzPN]{\includegraphics[width=0.15\textwidth]{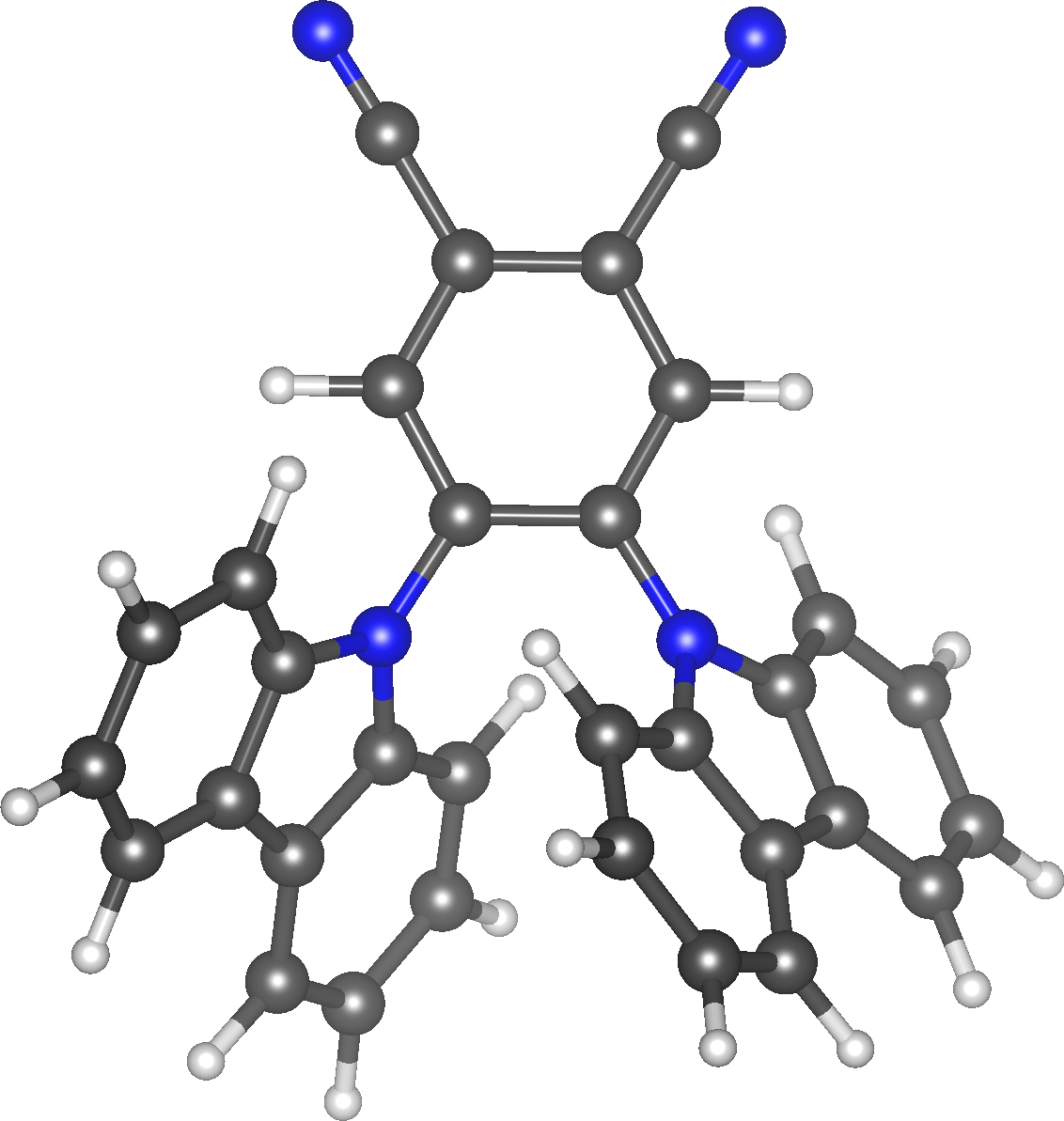}\label{fig:2czpn}}
\hspace{2pt}
\subfigure[ACRFLCN]{\includegraphics[width=0.15\textwidth]{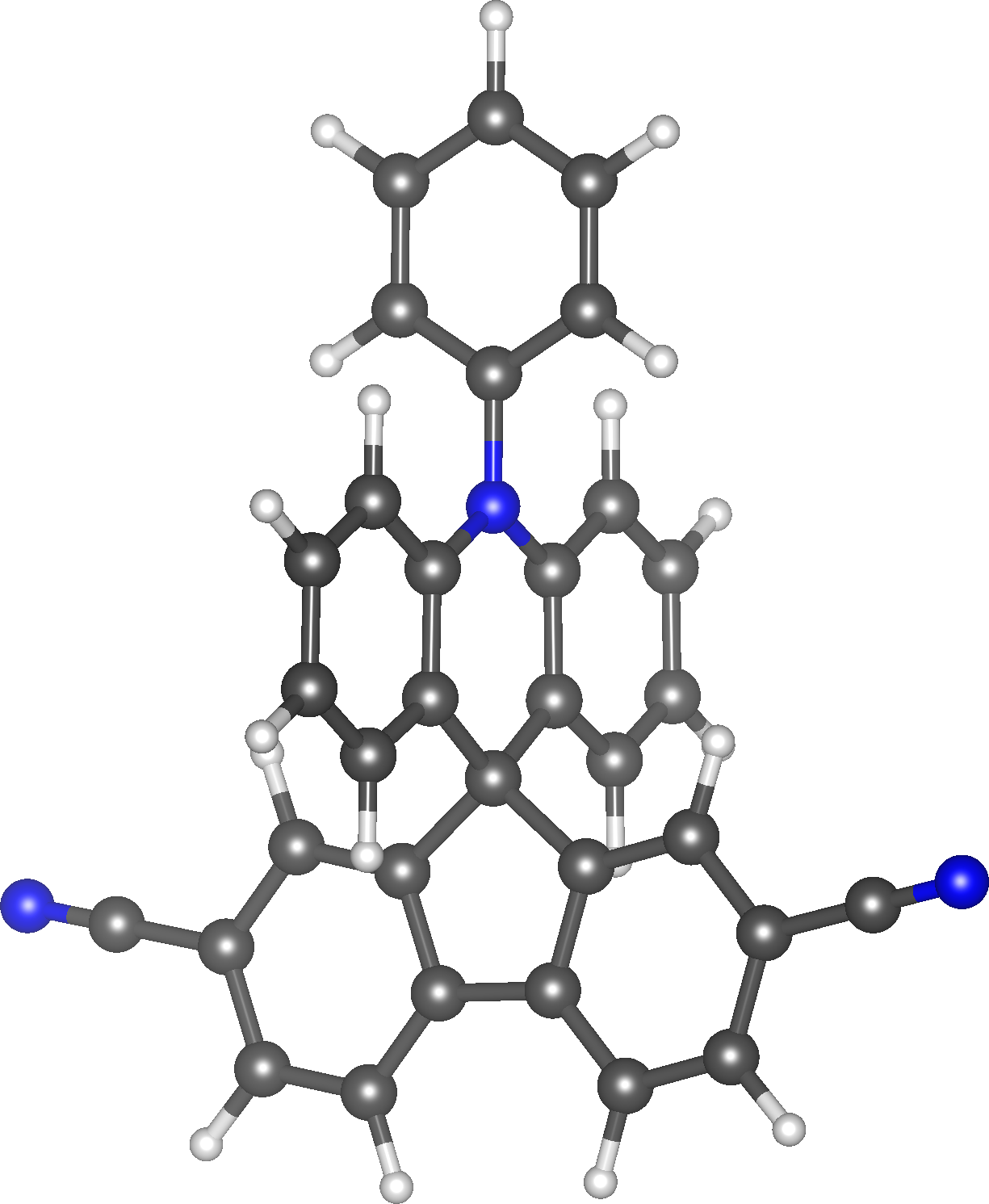}\label{fig:acrflcn}} 
        
\subfigure[CBP]{\includegraphics[width=0.22\textwidth]{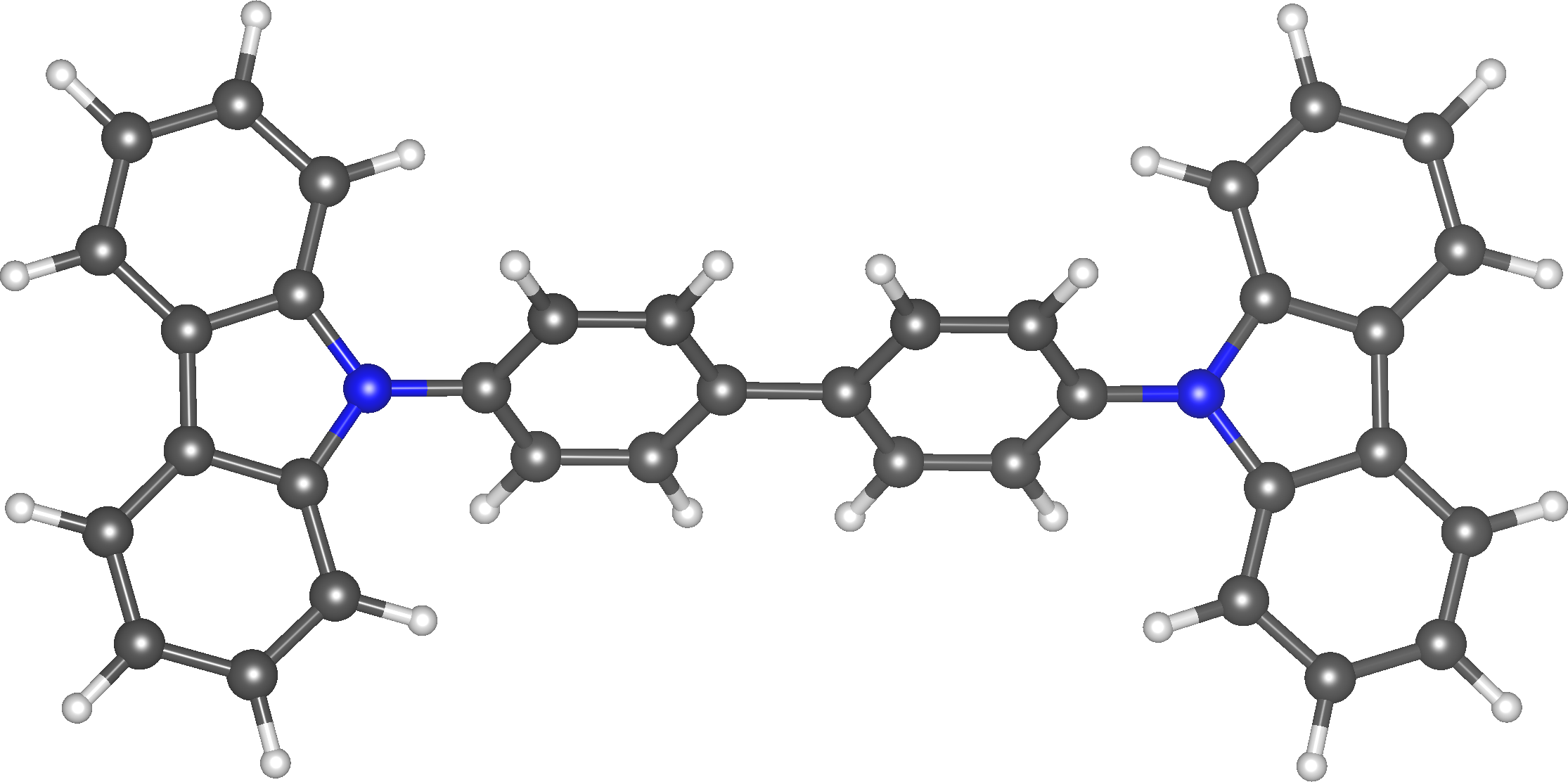}\label{fig:cbp}}

\caption{Relaxed atomic structures of the test set molecules. H/C/N atoms are depicted in white/black/blue. \label{fig:mols}}
\end{figure}

Benchmark calculations of the lowest energy singlet and triplet ($S_1$ and $T_1$) excitations were performed for a set of molecules which were chosen to provide a range of  LE, CT and mixed LE-CT character excitations. The test set, which is depicted in Fig.~\ref{fig:mols}, consists of five acenes, namely naphthalene, anthracene, tetracene, pentacene and hexacene, which constitute a set of well characterized molecules; and four exemplar OLED molecules, namely
NPh$_3$ (triphenylamine), 2CzPN (1,2-bis(carbazol-9-yl)-4,5-dicyanobenzene), ACRFLCN (10-phenyl-10H-spiro(acridine-9,9-
fluorene)-2,7-dicarbonitrile) and CBP (4,4'-Bis(N-carbazolyl)-1,1'-biphenyl).  NPh$_3$, 2CzPN and ACRFLCN are among the most investigated TADF emitters while CBP is a host molecule used to sensitize TADF emitters~\cite{Yang2017}. 

While experimental data for these systems is available, these are typically adiabatic excitation energies, which can differ significantly from vertical excitation energies (see e.g.\ Ref.~\cite{Hait2016}).
Furthermore, such experiments take place in various external conditions, with the results being sensitive to the external environment (i.e.\ the solvent or other molecular environment) as well as temperature~\cite{Huang2013,Yang2016}.
As the aim of this work is to assess the performance of T-CDFT for vertical excitations in gas phase, it is therefore not informative to make quantitative comparisons with experimental data. Indeed, one of the motivations behind this work is to provide a formalism which can treat large enough systems to take into account explicit environmental effects. Such comparisons are therefore saved for future work, while in the following we focus on theoretical comparisons only.

\subsection{Nature of the Excitations}

\begin{figure*}[ht]
\includegraphics[scale=0.5]{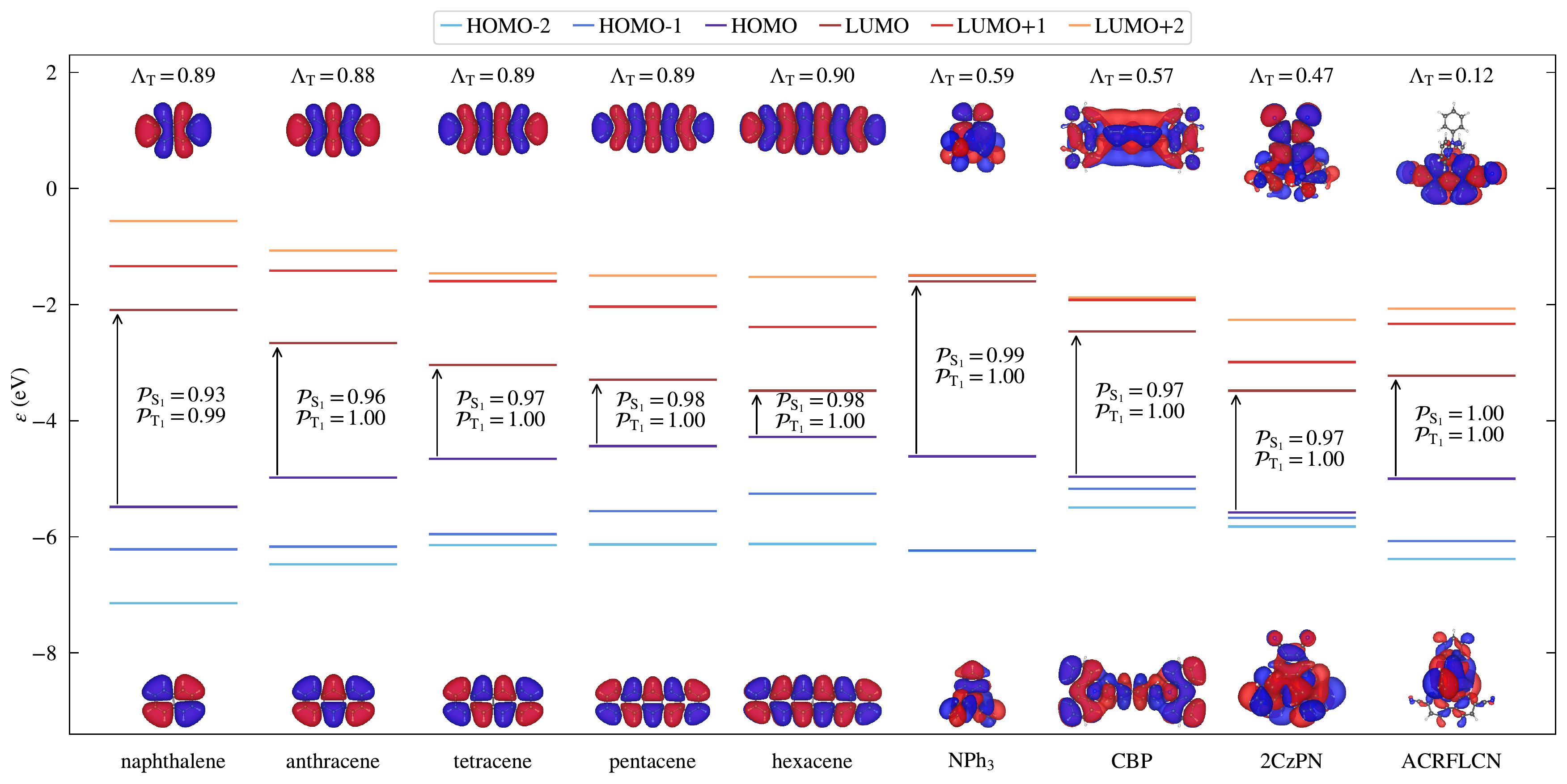}
\caption{PBE-calculated frontier orbital energies and corresponding HOMO and LUMO wavefunctions, as obtained from cubic scaling BigDFT. Wavefunctions were visualized in VESTA~\cite{Momma2011}, using an isosurface value of 0.0005~$a_0^{-3/2}$. The corresponding charge transfer parameter, $\Lambda_{\mathrm{T}}$ and HOMO-LUMO transition purity values, $\mathcal{P}$, are also given for each molecule, where the latter are calculated using LDA with a PBE basis, as described in the text.}
\label{fig:homo_lumo_pbe}
\end{figure*}

Before discussing the excitation energies, we first characterize the electronic excitations and component transitions for the benchmark molecules.  The frontier orbitals for PBE are depicted in Fig.~\ref{fig:homo_lumo_pbe}, alongside $\mathcal{P}$ and $\Lambda_{\mathrm{T}}$ values.
The equivalent PBE0 plot and the corresponding frontier orbital energies can be found in the Supporting Information.  

\subsubsection{Transition Purity}

The HOMO-LUMO transition purity values, $\mathcal{P}$, show that, within our computational set up, S$_{1}$ excitations are less pure than T$_{1}$, with the least pure excitation being 0.99 for T$_{1}$ and 0.93 for S$_{1}$.
We therefore expect that a HOMO-LUMO constraint in T-CDFT should represent a reasonable approximation for these molecules; in order to test this we treat the S$_{1}$ excitations of the acenes as both pure and mixed.  For the mixed excitations, the transition breakdown was taken from TDDFT, neglecting all contributions smaller than 0.01, and renormalizing the transition breakdown accordingly. The results are given in Table~\ref{tab:acene_mixed}. For naphthalene, which has the least pure excitation (0.93 before normalization), the energy for the mixed excitation is around 0.1~eV higher than the pure case. Whereas the mixed excitation energy for hexacene (0.98 purity before normalization), is only 0.02~eV higher than the pure excitation energy.  For other excitations which are much less pure than naphthalene, the mixed nature of the excitations may have a much stronger effect on the calculated energies, although this will also will vary depending on the involved transitions and not just the relative contributions. For the purposes of this work, however, these results suggest that a purity of 0.97-0.98 or above is such that neglecting other contributions should make little difference to the results.  Therefore, all OLED excitations are treated as pure, while the acene results in the following are those for the mixed excitations.

\begin{table*}[ht]
\centering
\begin{threeparttable}
\caption{Comparison of S$_1$ energies for the acenes calculated using T-CDFT with PBE both when treated as pure HOMO$\rightarrow$LUMO excitations, and when treated as mixed excitations including all transition contributions greater than 0.01. Shown are the normalized transition contributions, as well as the calculated energies in eV. \label{tab:acene_mixed}}
\begin{tabular*} {1.0\textwidth}{l @{\extracolsep{\fill}} rrrr}
\hline \hline
&HOMO$\rightarrow$  & HOMO-1$\rightarrow$ & HOMO-2$\rightarrow$ & \\
&LUMO &LUMO+1 & LUMO+2 & Energy\\
\cline{1-1}\cline{2-2}\cline{3-3}\cline{4-4}\cline{5-5}\\[-2.5ex]

\textbf{naphthalene}\\
pure & 1.000 & - & - & 			4.33\\
mixed & 0.935	& 0.027	& 0.038	& 4.41\\

\textbf{anthracene}\\

pure & 1.000 & - & - &			3.09\\
mixed & 0.975 &	0.010	& 0.015	& 3.15\\

\textbf{tetracene}\\
pure &	1.000	& - & - &	2.29\\
mixed & 0.988	& 0.012 & - &		2.32\\

\textbf{pentacene}\\
pure &	1.000	& - & - &	1.75\\
mixed & 0.989	& 0.011 & - &		1.77\\

\textbf{hexacene}\\
pure &	1.000	& - & - &	1.35\\
mixed & 0.989	& 0.011 & - &		1.37\\

 \hline \hline
\end{tabular*}
\end{threeparttable}
\end{table*}

\subsubsection{Spatial Overlap}

The high value of $\Lambda_{\mathrm{T}}$ for the acenes implies a strong spatial overlap between the HOMO and LUMO. Although this does not take into account the slightly mixed nature of the S$_1$ excitations in the shorter acenes, as a first approximation it implies that the transition constraint is local in nature, and will give rise to a predominantly local excitation (for both functionals). On the other hand, the OLED molecules have a smaller spatial overlap, so that the transition constraint and thus the excitation display a hybrid LE/CT nature of varying degrees, in agreement with previous results~\cite{Olivier2018}. We therefore expect TDDFT with PBE to perform more robustly for the acenes, whereas the CT character found in the OLED excitations could lead to a less accurate description.

\subsection{Acenes}

\begin{figure}[!h]
\centering
\includegraphics[scale=0.5]{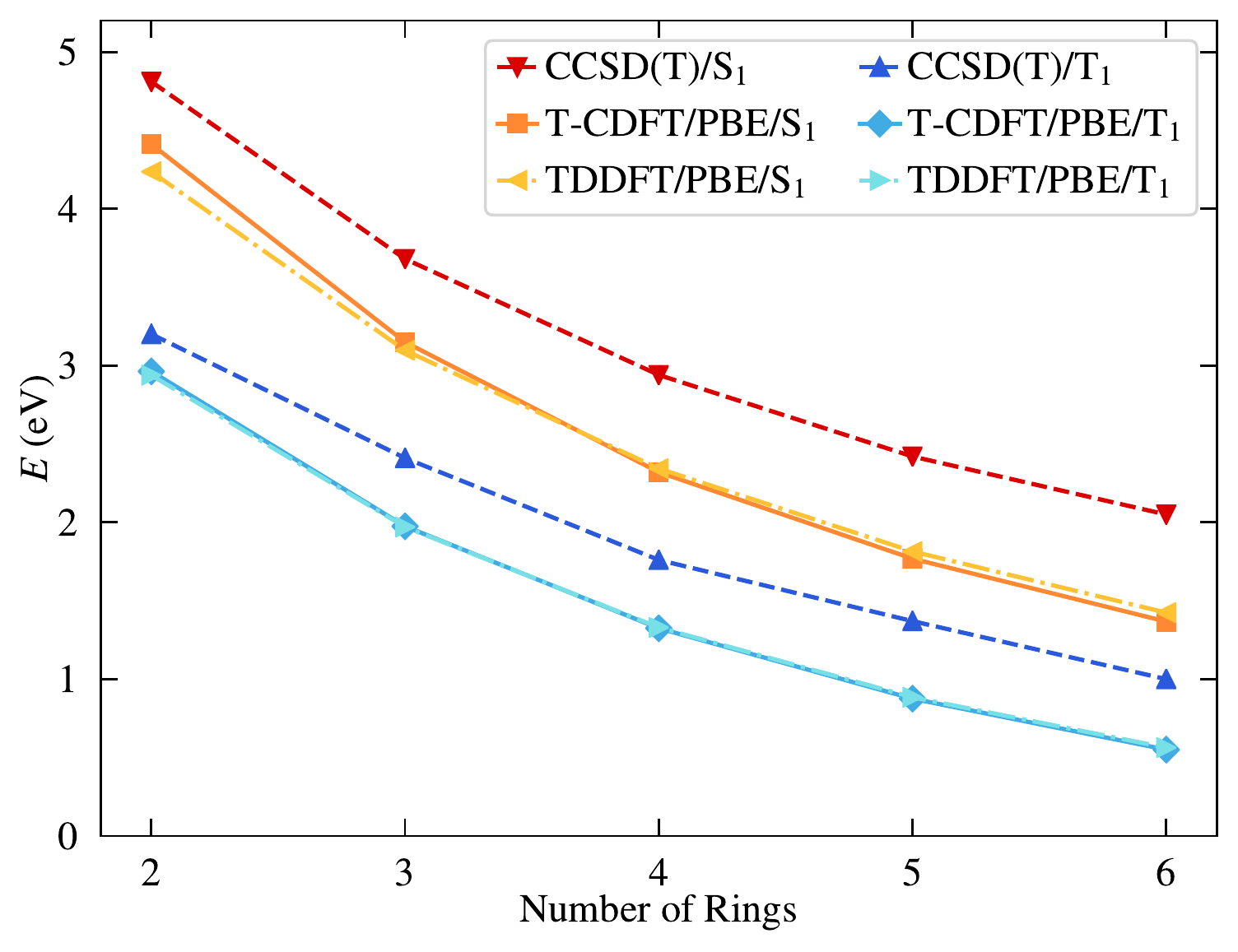}
\caption{Trend in S$_1$ and T$_1$ energies for the acenes from naphthalene to hexacene, for both T-CDFT and TDDFT with PBE, as well as CCSD(T), where the latter values are taken from Ref.~\cite{Rangel2017}. \label{fig:acene_trend}}
\end{figure}

We first consider the acenes, comparing our benchmark results with higher level theory calculations based on CCSD(T), which is often regarded as the gold standard of chemical accuracy in quantum chemistry~\cite{Bertels2021}. We employ the values from Ref~\cite{Rangel2017} (and references within). The low-lying singlet excitations in the acenes are termed $^1L_a$ and $^1L_b$, which differ in energetic ordering depending on the acene in question~\cite{Platt1949,Rangel2017}. However, since the $^1L_a$ state primarily involves a HOMO-LUMO transition and thus has the same character as our calculations, we take the $^1L_a$ values as our reference, irrespective of whether the CCSD(T)-calculated $^1L_b$ state is lower in energy. As shown in Fig.~\ref{fig:acene_trend}, both T-CDFT and TDDFT with PBE capture the CCSD(T) trend in S$_1$ and T$_1$ energies, albeit with a systematic underestimation of both states, which increases slightly with the number of rings. This underestimation is common to all the DFT-based approaches (see Supporting Information for tabulated results).

\begin{figure}[!h]
\centering\subfigure[Acenes]{\includegraphics[scale=0.5]{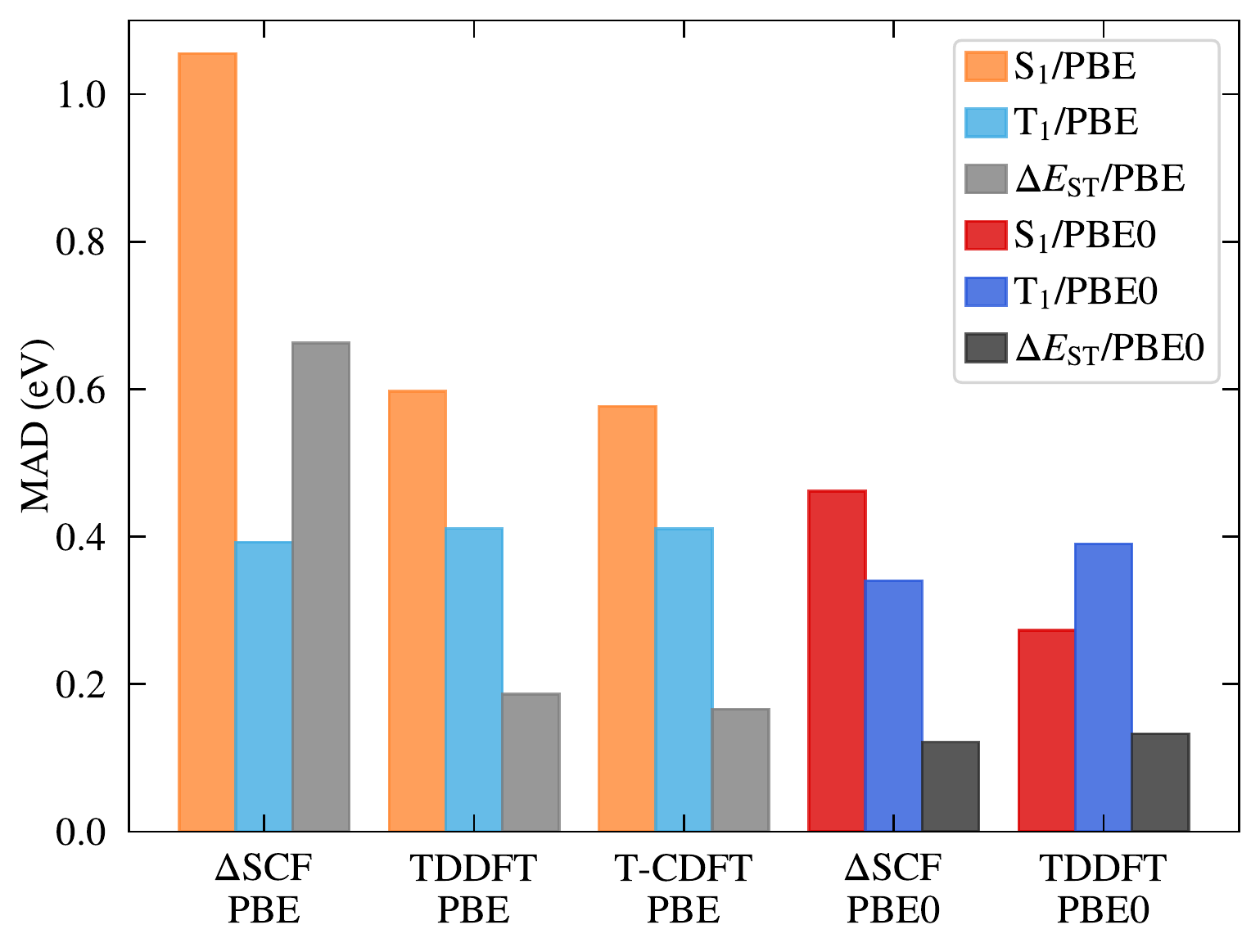}\label{fig:acenes_mad}}
\centering\subfigure[OLEDs]{\includegraphics[scale=0.5]{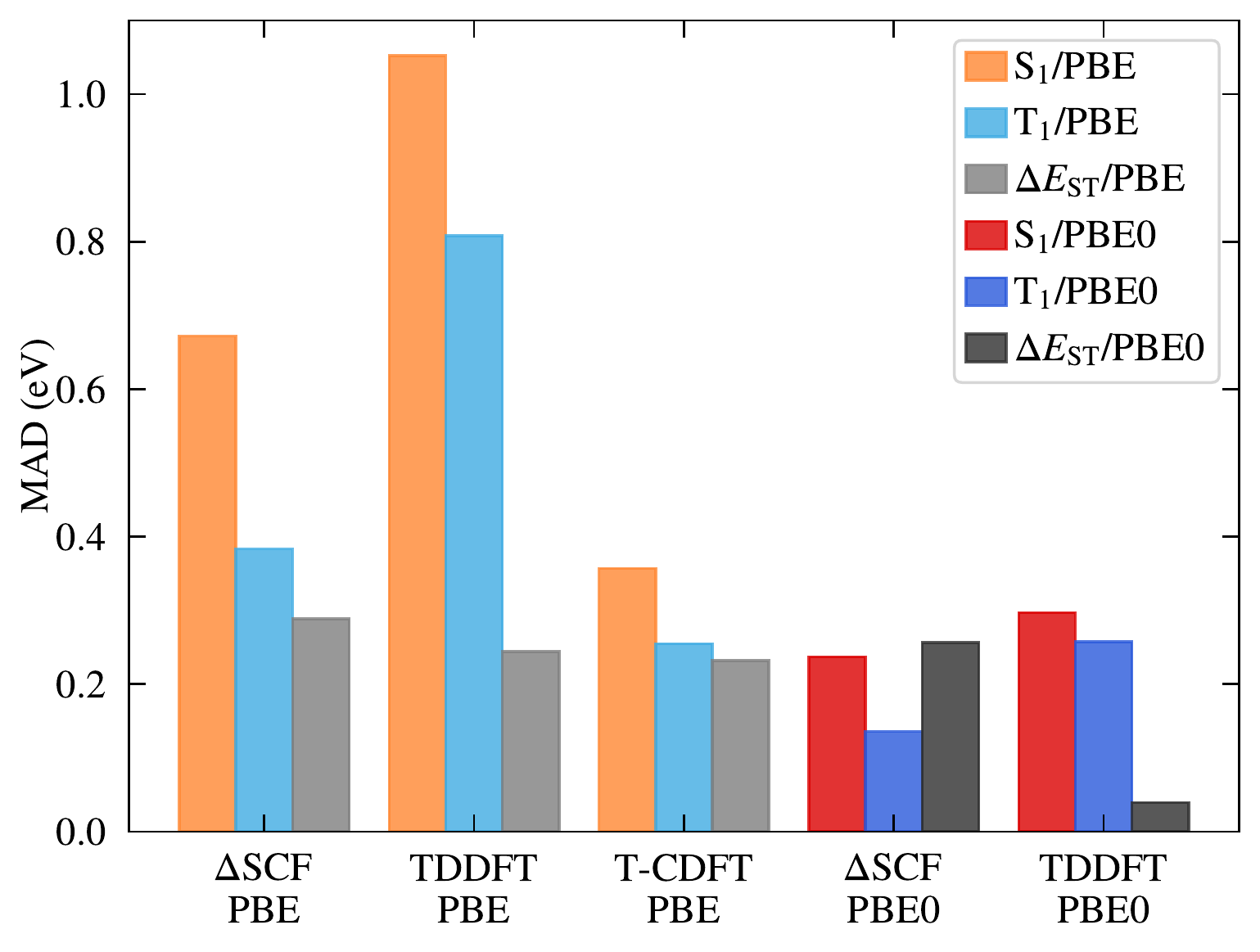}\label{fig:oleds_mad}}
\hspace{2pt}
\caption{Mean absolute deviation (MAD) of benchmark vertical S$_1$ and T$_1$ energies for the set of molecules exhibiting pure excitations, relative to reference energies coming from CCSD(T)~\cite{Rangel2017} for the acenes and TDA-TDDFT with a tuned range-separated functional~\cite{Sun2015} for the OLEDs. Corresponding energies are given in the Supporting Information.
\label{fig:mads}}
\end{figure}

Fig.~\ref{fig:acenes_mad} shows the mean absolute deviation (MAD) between each of the DFT-based approaches and CCSD(T). Compared to T$_1$, S$_1$ is more sensitive to both method and functional, with T$_1$ energies being relatively consistent across the benchmark results, and in most cases having a smaller MAD than S$_1$. Furthermore, the T-CDFT/PBE results are in remarkable agreement with TDDFT/PBE results despite the lower computational cost, with both approaches having MADs of 0.6 and 0.4~eV for S$_1$ and T$_1$ respectively.  Both T-CDFT and TDDFT significantly outperform $\Delta$SCF for S$_1$ when using PBE, while both $\Delta$SCF and TDDFT with PBE0 give a modest improvement in accuracy, albeit at much higher computational cost. In short, we find that T-CDFT with PBE performs very well for the predominantly local excitations seen in the acenes.

\subsection{OLEDs}

Due to their larger size, there is a lack of higher level quantum chemical reference data for the OLED molecules. However, due to the CT-like nature of the excitations, TDDFT with semi-local functionals cannot be expected to provide a reasonable reference.  Indeed, the limitations of TDDFT for CT states are well known, as is the corresponding strong dependence on the employed functional. This has motivated the use of range-separated hybrid functionals for the treatment of TADF materials.
For example, Adachi and co-workers~\cite{Huang2013} reported that functionals such as CAM-B3LYP~\cite{Yanai2004} or LC-$\omega$PBE~\cite{Scuseria2006} tend to overestimate absorption energies for common TADF molecules.
The situation may be improved by using ``optimally'' tuned range-separated functionals, which have been shown to give good agreement with experimental data~\cite{Penfold2015,Sun2015}, although the tuning of the separation parameters for a particular system increases the computational cost. Sun \emph{et al.}~\cite{Sun2015} computed vertical excitation energies for a set of OLED molecules, including those considered in this work, using TDA-TDDFT with an optimally tuned LC-$\omega$PBE* functional and a 6-31+G(d) basis set. We use these values as a reference in the following, although we note that they were performed using an implicit solvent (PCM toluene), which may influence the results.

\begin{figure}[!h]
\centering
\includegraphics[scale=0.5]{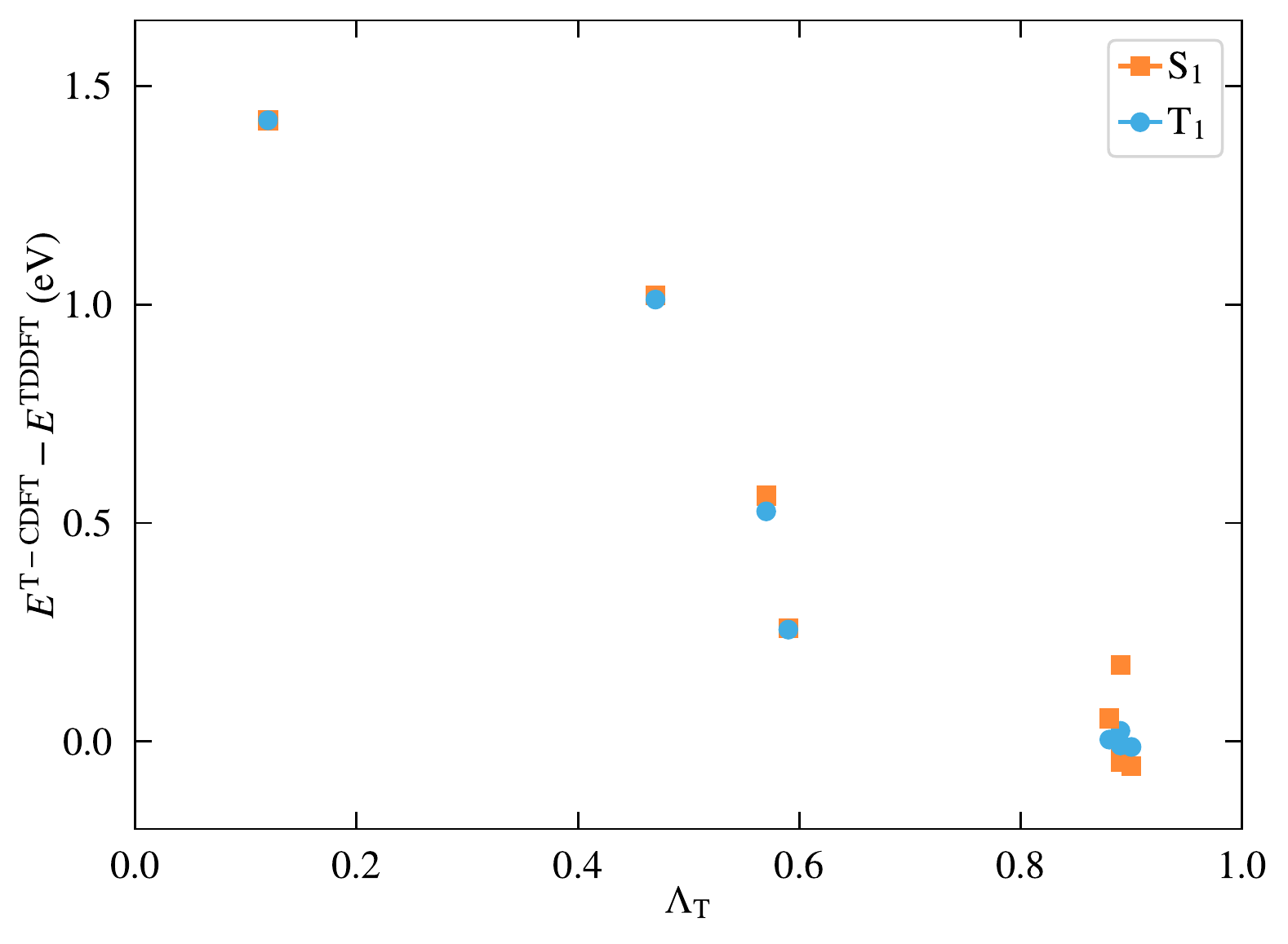}
\caption{Difference between T-CDFT and TDDFT energies \emph{vs}.\ $\Lambda_{\mathrm{T}}$, the HOMO-LUMO spatial overlap, where both calculations employ the PBE functional.\label{fig:ct_PBE}}
\end{figure}

The MAD between our calculations and the reference values is depicted in Fig.~\ref{fig:oleds_mad}.  There is a greater variability in MADs across methods and functionals compared to the acenes, particularly for T$_1$. Furthermore, both $\Delta$SCF and TDDFT with PBE systematically underestimate the reference values (see Supporting Information), with the large MAD for TDDFT/PBE being particularly striking. On the other hand, T-CDFT/PBE performs significantly better, giving MADs which are closer to the $\Delta$SCF and TDDFT PBE0 values. This much better performance of T-CDFT/PBE compared to TDDFT/PBE is in line with the more CT-like nature of the excitations.  Indeed, TDDFT/PBE most strongly underestimates the excitation energies for the molecules with the strongest CT-like character (i.e.\ the smallest $\Lambda_{\mathrm{T}}$ values).  At the same time, Fig.~\ref{fig:ct_PBE} shows that the smaller the value of $\Lambda_{\mathrm{T}}$, the bigger the difference between T-CDFT/PBE and TDDFT/PBE.  In other words, unlike TDDFT/PBE, which is strongly influenced by the nature of the transition, the quality of the T-CDFT/PBE results is not noticeably impacted by the nature of the excitation, giving reliable results for both the local excitations in acenes and the CT excitations in the OLED molecules.

Since accurate calculations of $\Delta E_{\mathrm{ST}}$ are crucial for designing new, optimal TADF emitters, we conclude this section by discussing $\Delta E_{\mathrm{ST}}$. Both $\Delta$SCF and TDDFT with PBE benefit to some degree from error cancellations in S$_1$ and T$_1$ errors, so that there is less variability across the methods. Nonetheless, while T-CDFT with PBE underestimates the reference $\Delta E_{\mathrm{ST}}$ values, this is less severe than the PBE-calculated $\Delta$SCF and TDDFT values. Furthermore, the MAD for T-CDFT/PBE is similar to that of $\Delta$SCF with PBE0, and only outperformed by the significantly more expensive TDDFT/PBE0 calculations.   

To sum up, what emerges from our benchmark calculations and other computational works~\cite{Moral2015,Sanz-Rodrigo2020,Olivier2018} is that the modelling of the excitations in OLED molecules is strongly method- and functional- dependant, and it is therefore not trivial to obtain a unique and consistent description. Nonetheless, by comparing with tuned range-separated functional calculations, we see that, unlike TDDFT with PBE, T-CDFT is equally able to model both LE and CT states.

\section{Conclusions}\label{sec:conclusions}

In this work we introduce a variation of CDFT (T-CDFT), wherein the constraint is defined as a transition between particular occupied and virtual orbitals, rather than a region of the simulation space as in traditional CDFT.
By defining an approach which goes beyond the linear response regime, we aim to provide a tool for the robust modelling of excitations in molecules. Our approach is applied to acenes and OLED emitters, for which the lowest energy singlet and triplet states are dominated by a transition between the HOMO and LUMO.  However, we also demonstrate the ability to take into account contributions from transitions between other orbitals. This has only a small impact on our benchmark calculations, but could prove to be important in future investigations of excitations with a more strongly mixed character.  

By comparing our benchmark calculations with reference values from the literature, we find that T-CDFT with PBE performs well for both the predominantly local excitations seen in the acenes, and the mix of CT and LE character seen in the OLED emitters, outperforming or equalling both $\Delta$SCF and TDDFT with the same functional. Importantly, T-CDFT does not suffer from the problems encountered when applying TDDFT with semi-local functionals to CT states, and, unlike CDFT with a spatial constraint, can model both LE and CT states. At the same time, the computational cost of T-CDFT is similar to the ground state, while the ability to use a fixed (large) Lagrange multipler keeps the cost significantly lower than TDDFT, even for mixed excitations involving multiple transitions.  Furthermore, T-CDFT also proves to be more robust than $\Delta$SCF, which can suffer from both spin contamination and convergence onto local minima.

Finally, our approach has been implemented in the linear scaling BigDFT code, and is fully compatible with the available fragmentation approaches. This capability could be used to impose excitations on a per-fragment basis in supramolecular or large biological systems. For example, in the case of local excitations on a molecule (fragment) in a given environment, where no strong coupling with the environment is expected, the constraint could be imposed between orbitals associated with the target fragment only, while still treating the \emph{full system}. Such an approach has the advantage of screening out spurious low energy charge transfer excitations which can be encountered with TDDFT.  On the other hand, in the case where charge transfer excitations between fragments are of interest, or where local excitations are expected to couple strongly with the environment, an alternative approach might be required. This could include performing TDDFT for a larger subset of the system, or using other information about the excitations to guide the choice of constraint(s).  Crucially, our framework is flexible enough to impose both intra- and inter-fragment constraints. 

In summary, T-CDFT provides a robust and accurate approach for treating both LE and CT states.  When combined with linear-scaling BigDFT, it is very well suited for treating excitations in large systems, enabling the exploration of explicit environmental effects on both excitation energies and $\Delta E_{\mathrm{ST}}$, a key quantity for modelling TADF-based OLEDs. 
Indeed, we foresee that such an approach will represent a powerful tool for the study of excitations in realistic supramolecular morphologies, for applications such as TADF.  Work in this direction is ongoing.

\begin{acknowledgement}

LER and MS acknowledge support from an EPSRC Early Career Research Fellowship (EP/P033253/1) and the Thomas Young Centre under grant number TYC-101. MS acknowledges Dr.\ Valerio Vitale for his initial help with automating the conversion between wavefunctions and support functions.
KT acknowledges support from the Engineering and Physical
Sciences Research Council (EP/S515085/1).
Calculations were performed on the Imperial College High Performance Computing Service, the ARCHER UK National Supercomputing Service, and the ARCHER2 UK National Supercomputing Service (https://www.archer2.ac.uk).

The Supporting Information contains additional computational details and further results. This includes a discussion of the investigation into local minima in $\Delta$SCF, an example plot showing the effect of varying the Lagrange multiplier, a plot of the frontier orbital energies as calculated with PBE0, and tabulated data for the frontier orbital energies calculated using different basis sets, and both a plot and table containing the excitation energies calculated using the different approaches. In addition, the data associated with this work, including Jupyter notebooks and associated files which can be used to reproduce the calculations, are available at \url{https://gitlab.com/martistella86/t-cdft-notebooks.}

\end{acknowledgement}


\bibliography{references}

\end{document}


\date{\today}

\author{Martina Stella}     
\affiliation{\ICL}
\affiliation{\ICTP}
\author{Kritam Thapa}  
\affiliation{\ICL}
\author{Luigi Genovese}
\affiliation{\CEA}
\author{Laura E.\ Ratcliff}
\affiliation{\ICL}
\affiliation{\Bristol}

\title{{\LARGE Supplementary Information}\\
\vspace{5pt}Transition-Based Constrained DFT for the Robust and Reliable Treatment of Excitations in Supramolecular Systems}

\maketitle

\subsection{Additional Computational Details}

BigDFT calculations were performed using HGH-GTH PSPs~\cite{Goedecker1996,Hartwigsen1998} with non-linear core corrections~\cite{Willand2013}.
Gas phase geometry optimizations were performed for the ground state using PBE in BigDFT, with a maximum force threshold of 0.02~eV/\AA.  Single point calculations were performed with a wavelet grid spacing of~0.26~\AA, while geometry optimizations employed a smaller grid spacing of 0.24~\AA\ to ensure accurate force calculations. All BigDFT calculations used coarse and fine multipliers of 7 and 9 respectively.  Cubic scaling BigDFT calculations employed a gradient convergence threshold of $10^{-5}$.  
Due to the poor convergence behaviour of some NWChem calculations, a strict upper limit of 200 iterations was imposed for both $\Delta$SCF and TDDFT calculations, beyond which calculations were considered not to have reached convergence.

T-CDFT WFN-based calculations were performed for a range of basis sizes, while SF-based calculations were performed for a range of localization radii for two basis set sizes -- a minimal basis with 1/4/4 basis functions per H/C/N atom and a larger basis with 4/9/9 basis functions per H/C/N atom. The minimal basis was insufficient, giving energies almost 0.2~eV higher in energy than the more converged basis. On the other hand, increasing the localization radius from 4.23 to 5.29~\AA\ led to differences less than 0.05~eV.  We note that, due to the presence of additional degrees of freedom coming from environment molecules, it may be possible to use a smaller SF basis while retaining the same accuracy in future T-CDFT calculations of larger systems. 
All SF-based T-CDFT calculations were performed for a basis set optimized to represent all negative energy (bound) virtual states, as identified from the equivalent cubic scaling PBE calculation.  No kernel truncation was applied, since the addition of the constraint decreases the locality of the kernel.  

The HOMO-LUMO spatial overlap, $\Lambda_{\mathrm{T}}$, was calculated as a post-processing calculation using the HOMO and LUMO wavefunctions extracted from cubic-scaling BigDFT. The HOMO-LUMO transition purity, $\mathcal{P}$, or the full transition breakdown in the case of mixed excitations, was also calculated as a post-processing step following BigDFT TDDFT calculations with LDA in a PBE-generated WFN basis, for which care was also taken to ensure the basis set contained enough virtual states to reach convergence.  Although not used in this work, the functionality also exists to calculate transition purities using a SF basis.

\subsection{Local Minima in $\Delta$SCF}

For calculations which were performed using both BigDFT and NWChem, a close comparison of results was performed. In general, the results showed good agreement, however there were some large deviations between $\Delta$SCF values between the two codes, which could not be attributed to basis set differences or the use of pseudopotentials, even when accounting for the fact that purification increases uncertainties in S$_1$ coming from basis set, PSP and convergence differences.  Indeed, although generally considered a reliable tool for the computation of vertical HOMO-LUMO transitions in organic emitters~\cite{Kowalczyk2011}, because SCF algorithms are geared toward energy minimization, $\Delta$SCF is known to sometimes collapse to low energy states. This can include the ground state~\cite{Bourne2021}, although methods such as the maximum overlap method~\cite{Gilbert2008} have been developed to alleviate this problem. For some OLED molecules, it has also been observed to converge onto a different excited state due to a poor orbital guess or degeneracy~\cite{Zhao2021}. 

In this work no collapses onto the ground state were observed. However, large initial variations between BigDFT and NWChem energies were found, which could not be explained by basis set differences, even taking into account the fact that the use of the purification formula can exaggerate differences in singlet energies if the triplet states also differ. Instead, the variations were attributed to the presence of local minima, which was confirmed by inspecting the charge density differences between the ground state and the corresponding excited state.
It was found that using the ground state orbitals as an initial guess for excited state calculations eliminated these local minima, typically also improving convergence. This was confirmed by again inspecting the charge density differences.

\subsection{Lagrange Multiplier}

\begin{figure*}[!h]
\centering
\includegraphics[scale=0.5]{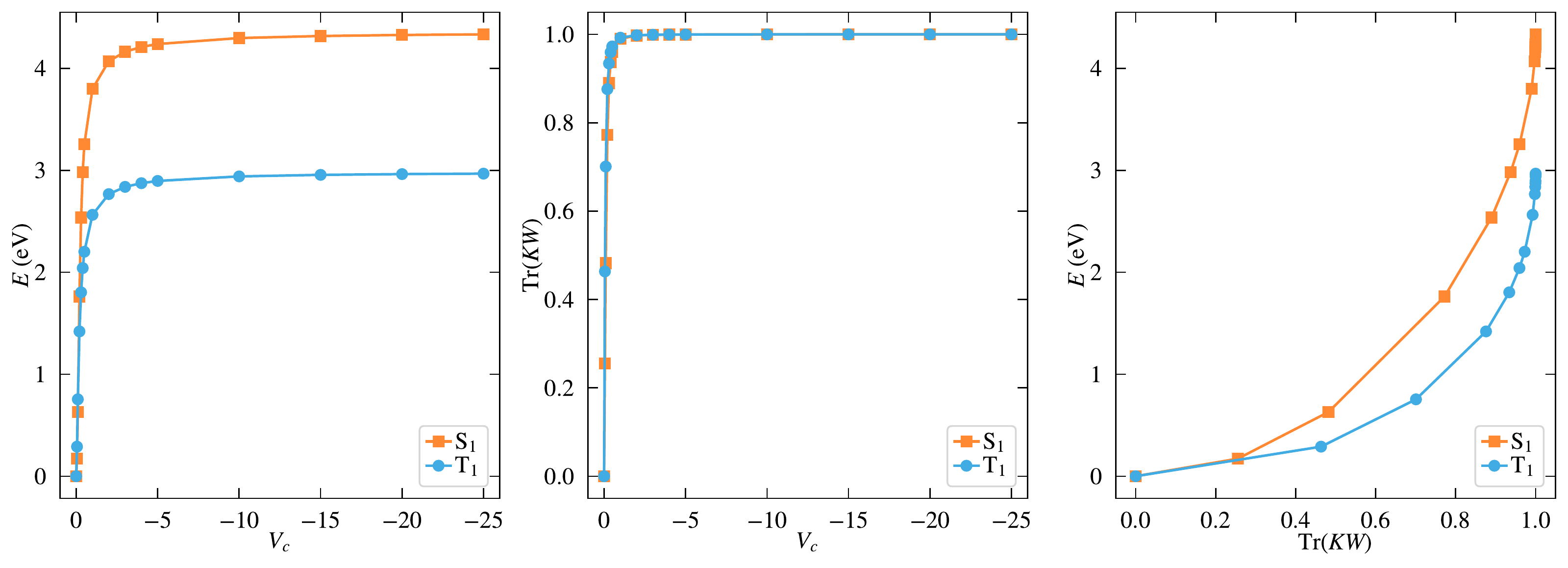}
\caption{Influence of the Lagrange multiplier, $V_c$, on the T-CDFT calculated singlet and triplet energies for a pure HOMO-LUMO constraint in naphthalene, as well as the constrained charge, $\mathrm{Tr}(KW)$, and the relation between the two. Calculations are performed for a SF basis with 4(9) SFs per H(C) atom, with localization radii $R_{\mathrm{loc}}=4.23$~\AA. \label{fig:naphthalene_vc_convergence}}
\end{figure*}

Fig.~\ref{fig:naphthalene_vc_convergence} shows the effect of varying the Lagrange multiplier for a pure excitation in naphthalene. A value of -20 was used in all calculations, giving $\mathrm{Tr}\left(\mathbf{KW}\right)$ within 0.01 of 1$e^-$ and thus confirming that there is no need to optimize $V_c$ on a case-by-case basis.

\clearpage

\subsection{Additional Results}

\begin{figure*}[!h]
\includegraphics[scale=0.5]{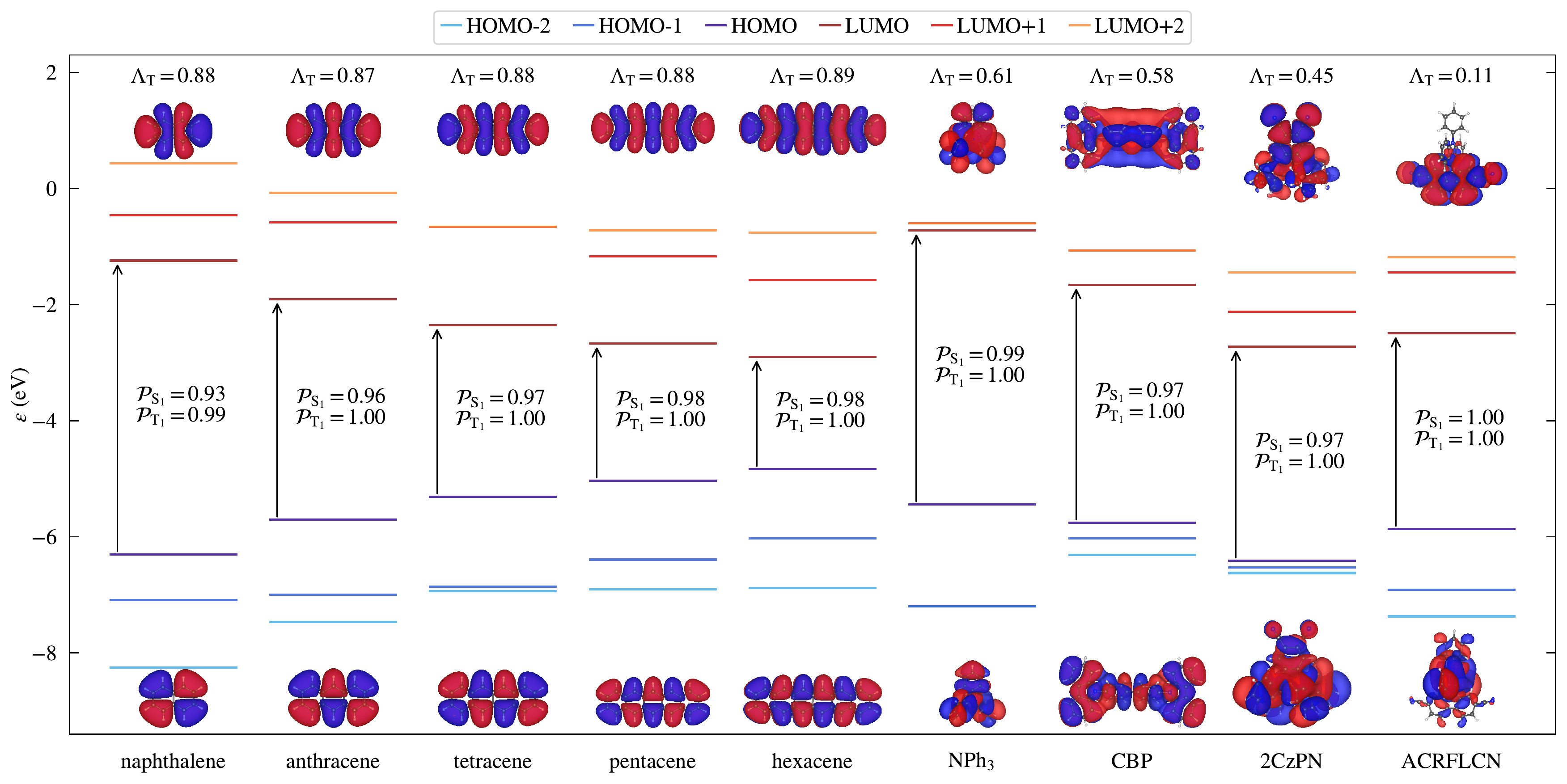}
\caption{PBE0-calculated frontier orbital energies and corresponding HOMO and LUMO wavefunctions, as obtained from NWChem using the cc-pVTZ basis. Wavefunctions were visualized in VESTA, using an isosurface value of 0.0005~$a_0^{-3/2}$. The corresponding charge transfer parameter, $\Lambda_{\mathrm{T}}$ and HOMO-LUMO transition purity values, $\mathcal{P}$, are also given for each molecule, where the latter are calculated using LDA with a PBE basis, as described in the main text. The corresponding energies are given in Tables~\ref{tab:acene_orbital_energies} and~\ref{tab:oled_orbital_energies}. \label{fig:homo_lumo_pbe0}}
\end{figure*}

\clearpage

\begin{table*}[ht]
\centering
\begin{threeparttable}
\caption{Frontier orbital energies and band gaps of the acenes, calculated using different functionals with the cc-pVTZ (`VTZ') basis set in NWChem, a wavelet basis using the cubic scaling approach of BigDFT and a SF basis of 4/9/9 SFs per H/C/N atom with $R_{\mathrm{loc}}=4.23$~\AA. All values are in eV. \label{tab:acene_orbital_energies}}

\begin{tabular*} {1.0\textwidth}{l @{\extracolsep{\fill}} rrr rr}
\hline \hline
& \multicolumn{3}{c}{PBE} & \multicolumn{2}{c}{PBE0}\\
  \cline{2-4}\cline{5-6}\\[-2.5ex]
  & VTZ & wavelet &  SF & VTZ & wavelet   \\
\cline{1-1}\cline{2-2}\cline{3-3}\cline{4-4}\cline{5-5}\cline{6-6}\\[-2.5ex]

\textbf{naphthalene}\\
HOMO-2 &  -7.11 &   -7.15 &  -7.16 &  -8.23 &   -8.25  \\
HOMO-1 &  -6.18 &   -6.22 &  -6.23 &  -7.07 &   -7.09  \\
HOMO &  -5.45 &   -5.48 &  -5.50 &  -6.30 &   -6.30 \\
LUMO &  -2.04 &   -2.09 &  -2.09 &  -1.22 &   -1.24 \\
LUMO+1 &  -1.28 &   -1.34 &  -1.30 &  -0.43 &   -0.46  \\
LUMO+2 &  -0.43 &   -0.56 &  -0.45 &   0.59 &    0.43  \\
gap &   3.40 &    3.39 &   3.41 &   5.07 &    5.06  \\
\\
\textbf{anthracene}\\
HOMO-2 &  -6.44 &   -6.47 &  -6.48 &  -7.46 &   -7.47  \\
HOMO-1 &  -6.14 &   -6.17 &  -6.18 &  -6.98 &   -6.99  \\
HOMO &  -4.95 &   -4.98 &  -5.00 &  -5.70 &   -5.71 \\
LUMO &  -2.63 &   -2.67 &  -2.67 &  -1.90 &   -1.91 \\
LUMO+1 &  -1.37 &   -1.42 &  -1.38 &  -0.57 &   -0.58  \\
LUMO+2 &  -1.00 &   -1.07 &  -1.03 &  -0.03 &   -0.08  \\
gap &   2.32 &    2.32 &   2.33 &   3.80 &    3.80  \\
\\
\textbf{tetracene}\\
HOMO-2 &  -6.11 &   -6.15 &  -6.16 &  -6.93 &   -6.94  \\
HOMO-1 &  -5.92 &   -5.95 &  -5.97 &  -6.86 &   -6.86  \\
HOMO &  -4.63 &   -4.66 &  -4.67 &  -5.31 &   -5.31  \\
LUMO &  -3.00 &   -3.04 &  -3.05 &  -2.35 &   -2.35 \\
LUMO+1 &  -1.55 &   -1.60 &  -1.59 &  -0.65 &   -0.66 \\
LUMO+2 &  -1.42 &   -1.46 &  -1.46 &  -0.65 &   -0.66  \\
gap &   1.62 &    1.62 &   1.62 &   2.96 &    2.96  \\
\\
\textbf{pentacene}\\
HOMO-2 &  -6.10 &   -6.13 &  -6.15 &  -6.90 &   -6.90  \\
HOMO-1 &  -5.53 &   -5.56 &  -5.58 &  -6.39 &   -6.39  \\
HOMO &  -4.41 &   -4.44 &  -4.45 &  -5.04 &   -5.04  \\
LUMO &  -3.27 &   -3.30 &  -3.31 &  -2.67 &   -2.67  \\
LUMO+1 &  -2.00 &   -2.03 &  -2.04 &  -1.17 &   -1.17  \\
LUMO+2 &  -1.46 &   -1.50 &  -1.49 &  -0.72 &   -0.72  \\
gap &   1.14 &    1.14 &   1.14 &   2.37 &    2.37 \\
\\
\textbf{hexacene}\\
HOMO-2 &  -6.10 &   -6.12 &  -6.14 &  -6.88 &   -6.88  \\
HOMO-1 &  -5.23 &   -5.26 &  -5.27 &  -6.03 &   -6.03  \\
HOMO &  -4.25 &   -4.28 &  -4.29 &  -4.85 &   -4.84  \\
LUMO &  -3.46 &   -3.48 &  -3.49 &  -2.90 &   -2.90  \\
LUMO+1 &  -2.35 &   -2.38 &  -2.39 &  -1.58 &   -1.58  \\
LUMO+2 &  -1.49 &   -1.52 &  -1.51 &  -0.76 &   -0.76  \\
gap &   0.79 &    0.80 &   0.80 &   1.94 &    1.93 \\

 \hline \hline
\end{tabular*}
\end{threeparttable}
\end{table*}

\begin{table*}
\centering
\begin{threeparttable}
\caption{Frontier orbital energies and band gaps of the OLED molecules, calculated using different functionals with the cc-pVTZ (`VTZ') basis set in NWChem, a wavelet basis using the cubic scaling approach of BigDFT and a SF basis of 4/9/9 SFs per H/C/N atom with $R_{\mathrm{loc}}=4.23$~\AA. All values are in eV. \label{tab:oled_orbital_energies}}

\begin{tabular*} {1.0\textwidth}{l @{\extracolsep{\fill}} rrr rr}
\hline \hline
& \multicolumn{3}{c}{PBE} & \multicolumn{2}{c}{PBE0}\\
  \cline{2-4}\cline{5-6}\\[-2.5ex]
  & VTZ & wavelet &  SF & VTZ & wavelet   \\
\cline{1-1}\cline{2-2}\cline{3-3}\cline{4-4}\cline{5-5}\cline{6-6}\\[-2.5ex]

\textbf{NPh3}\\
HOMO-2 &  -6.20 &   -6.24 &  -6.28 &  -7.18 &   -7.20  \\
HOMO-1 &  -6.20 &   -6.24 &  -6.28 &  -7.18 &   -7.20  \\
HOMO &  -4.57 &   -4.61 &  -4.65 &  -5.43 &   -5.45 \\
LUMO &  -1.53 &   -1.60 &  -1.60 &  -0.67 &   -0.72  \\
LUMO+1 &  -1.43 &   -1.50 &  -1.51 &  -0.55 &   -0.60  \\
LUMO+2 &  -1.43 &   -1.49 &  -1.51 &  -0.55 &   -0.60  \\
gap &   3.04 &    3.01 &   3.05 &   4.77 &    4.72  \\
\\
\textbf{CBP}\\
HOMO-2 &  -5.46 &   -5.50 &  -5.52 &  -6.31 &   -6.31  \\
HOMO-1 &  -5.14 &   -5.17 &  -5.20 &  -6.01 &   -6.03  \\
HOMO &  -4.92 &   -4.96 &  -4.99 &  -5.75 &   -5.76 \\
LUMO &  -2.41 &   -2.46 &  -2.47 &  -1.65 &   -1.66 \\
LUMO+1 &  -1.86 &   -1.92 &  -1.91 &  -1.04 &   -1.07 \\
LUMO+2 &  -1.82 &   -1.88 &  -1.88 &  -1.04 &   -1.07 \\
gap &   2.51 &    2.50 &   2.52 &   4.11 &    4.10 \\
\\
\textbf{2CzPN}\\
HOMO-2 &  -5.78 &   -5.82 &  -5.85 &  -6.59 &   -6.62 \\
HOMO-1 &  -5.63 &   -5.67 &  -5.70 &  -6.49 &   -6.53  \\
HOMO &  -5.54 &   -5.58 &  -5.61 &  -6.38 &   -6.41  \\
LUMO &  -3.43 &   -3.48 &  -3.50 &  -2.70 &   -2.73 \\
LUMO+1 &  -2.93 &   -2.99 &  -3.01 &  -2.08 &   -2.12  \\
LUMO+2 &  -2.20 &   -2.26 &  -2.27 &  -1.39 &   -1.44 \\
gap &   2.11 &    2.10 &   2.11 &   3.68 &    3.69  \\
\\
\textbf{ACRFLCN}\\
HOMO-2 &  -6.34 &   -6.38 &  -6.40 &  -7.35 &   -7.37 \\
HOMO-1 &  -6.03 &   -6.07 &  -6.09 &  -6.90 &   -6.91 \\
HOMO &  -4.95 &   -5.00 &  -5.02 &  -5.85 &   -5.87 \\
LUMO &  -3.17 &   -3.22 &  -3.23 &  -2.48 &   -2.49 \\
LUMO+1 &  -2.26 &   -2.33 &  -2.32 &  -1.40 &   -1.45 \\
LUMO+2 &  -2.01 &   -2.07 &  -2.07 &  -1.16 &   -1.19 \\
gap &   1.79 &    1.78 &   1.79 &   3.37 &    3.37  \\

 \hline \hline
\end{tabular*}
\end{threeparttable}
\end{table*}

\begin{figure*}[ht]
\centering
\includegraphics[scale=0.5]{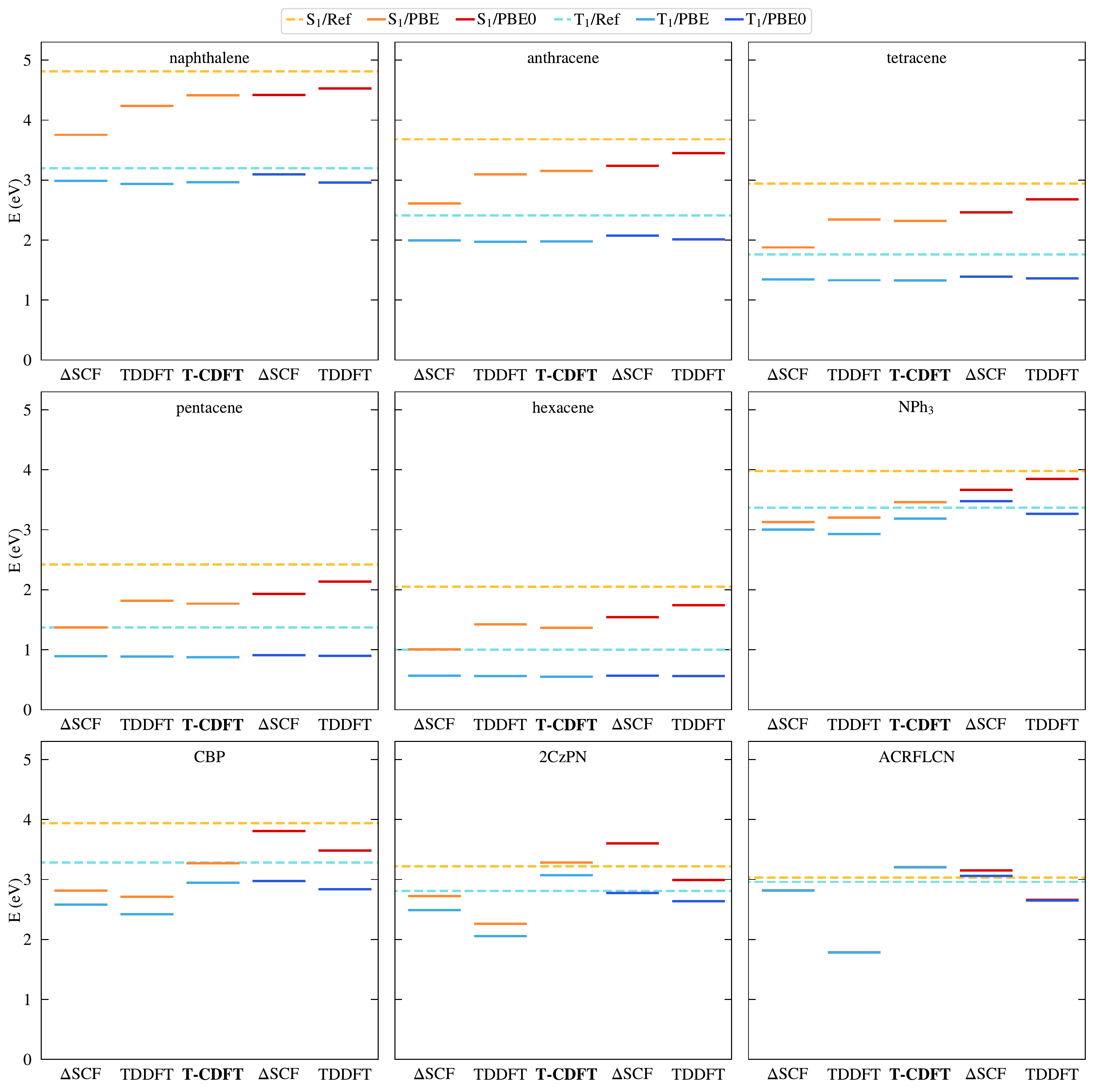}
\caption{Vertical S$_1$ and T$_1$ energies computed using different methods (T-CDFT, $\Delta$SCF, TDDFT) and functionals (PBE, PBE0) using BigDFT and NWChem.
Corresponding energies are given in Table~\ref{tab:excitations}, while the reference energies are calculated using CCSD(T)~\cite{Rangel2017, Lopata2011} for the acenes, and TDA-TDDFT with a tuned range-separated functional~\cite{Sun2015} for the OLEDs.\label{fig:excitations}}
\end{figure*}

\clearpage

\begin{table*}
\centering
\begin{threeparttable}
\caption{Vertical S$_{1}$ and T$_{1}$ energies and singlet-triplet splittings, $\Delta E_{\mathrm{ST}}$, in eV, calculated using different methods (T-CDFT, $\Delta$SCF, TDDFT) and functionals (PBE, PBE0) using BigDFT and NWChem, as compared to reference values from CCSD(T)~\cite{Rangel2017, Lopata2011} for the acenes, and TDA-TDDFT with a tuned range-separated functional~\cite{Sun2015} for the OLEDs.  Also given is the HOMO-LUMO transition purity, $\mathcal{P}$, the spatial overlap, $\Lambda_T$, and the mean absolute deviation (MAD) of the different quantities with respect to the reference values. }
\label{tab:excitations}
\begin{tabular*} {1.0\textwidth}{l @{\extracolsep{\fill}} r r rrrr rrr}
\hline \hline
& & & \multicolumn{4}{c}{PBE}  & \multicolumn{3}{c}{PBE0} \\
 \cline{4-7}\cline{8-10}\\[-2.5ex]
& Ref & $\mathcal{P}$ & $\Lambda_{\mathrm{T}}$ & $\Delta$SCF & TDDFT & T-CDFT & $\Lambda_{\mathrm{T}}$ &  $\Delta$SCF & TDDFT\\
 \cline{1-1}\cline{2-2}\cline{3-3}\cline{4-4}\cline{5-5}\cline{6-6}\cline{7-7}\cline{8-8}\cline{9-9}\cline{10-10}\\[-2.5ex]

\textbf{naphthalene}\\
$S_\mathrm{1}$ & 4.81 &  0.93 &     &  3.75 &  4.24 &  4.41 &             &  4.42 &  4.53 \\
$T_\mathrm{1}$ & 3.20 &  0.99 &   0.89 &  2.98 &  2.94 &  2.96 &        0.88 &  3.10 &  2.96 \\
$\Delta E_{\mathrm{ST}}$ &  1.61 & &   &  0.77 &  1.30 &  1.36 &             &  1.32 &  1.57 \\

\textbf{anthracene}\\
$S_\mathrm{1}$ & 3.68 &  0.96 &           &  2.61 &  3.10 &  3.15 &             &  3.24 &  3.45 \\
$T_\mathrm{1}$ & 2.41 &  1.00 &   0.88 &  1.99 &  1.97 &  1.97 &        0.87 &  2.07 &  2.01 \\
$\Delta E_{\mathrm{ST}}$ &  1.27 &  &  &  0.62 &  1.13 &  1.12 &             &  1.16 &  1.44 \\

\textbf{tetracene}\\
$S_\mathrm{1}$ & 2.94 & 0.97 &       &  1.88 &  2.34 &  2.32 &             &  2.46 &  2.68 \\
$T_\mathrm{1}$ & 1.76 & 1.00 &     0.89 &  1.34 &  1.33 &  1.32 &        0.88 &  1.39 &  1.36 \\
$\Delta E_{\mathrm{ST}}$ &  1.18 &  &    &  0.54 &  1.01 &  0.97 &             &  1.07 &  1.32 \\

\textbf{pentacene}\\
$S_\mathrm{1}$ & 2.42 &  0.98 &    &  1.37 &  1.81 &  1.75 &             &  1.93 &  2.14 \\
$T_\mathrm{1}$ & 1.37 &  1.00 &  0.89 &  0.89 &  0.88 &  0.88 &        0.88 &  0.91 &  0.90 \\
$\Delta E_{\mathrm{ST}}$ &  1.05 &  &      &  0.48 &  0.93 &  0.87 &             &  1.02 &  1.24 \\

\textbf{hexacene}\\
$S_\mathrm{1}$ & 2.05 &  0.98 &        &  1.01 &  1.42 &  1.37 &             &  1.55 &  1.74 \\
$T_\mathrm{1}$ & 1.00 &  1.00 &    0.90 &  0.57 &  0.56 &  0.55 &        0.89 &  0.56 &  0.56 \\
$\Delta E_{\mathrm{ST}}$ & 1.05 &  &    &  0.44 &  0.86 &  0.80 &             &  0.98 &  1.18 \\

\textbf{MAD}\\
$S_\mathrm{1}$ &  &  &  & 1.06 &  0.60 &  0.58 &            &  0.46 &  0.27 \\
$T_\mathrm{1}$ &  &  &  & 0.39 &  0.41 &  0.41 &           &  0.34 &  0.39 \\
$\Delta E_{\mathrm{ST}}$ &   &   &  &  0.66 &  0.19 &  0.17 &            &  0.12 &  0.13 \\

 \cline{1-1}\cline{2-2}\cline{3-3}\cline{4-4}\cline{5-5}\cline{6-6}\cline{7-7}\cline{8-8}\cline{9-9}\cline{10-10}\\[-2.5ex]
\textbf{NPh$_3$}\\
$S_\mathrm{1}$ & 3.98 & 0.99 &     &  3.13 &  3.20 &  3.46 &             &  3.67 &  3.85 \\
$T_\mathrm{1}$ & 3.37 & 1.00 &    0.59 &  3.00 &  2.93 &  3.19 &        0.61 &  3.47 &  3.27 \\
$\Delta E_{\mathrm{ST}}$ & 0.61 &  &     &  0.13 &  0.27 &  0.28 &             &  0.19 &  0.58 \\

\textbf{CBP}\\
$S_\mathrm{1}$ & 3.94 &  0.97 &        &  2.81 &  2.71 &  3.27 &             &  3.81 &  3.48 \\
$T_\mathrm{1}$ & 3.28 &  1.00 &     0.57 &  2.58 &  2.42 &  2.95 &        0.58 &  2.98 &  2.84 \\
$\Delta E_{\mathrm{ST}}$ & 0.66 &  &   &  0.23 &  0.29 &  0.33 &             &  0.83 &  0.64 \\
    
\textbf{2CzPN}\\
$S_\mathrm{1}$ & 3.22 &  0.97 &           &  2.72 &  2.26 &  3.28 &             &  3.60 &  2.99 \\
$T_\mathrm{1}$ & 2.81 &  1.00 &    0.47 &  2.49 &  2.06 &  3.07 &        0.45 &  2.78 &  2.64 \\
$\Delta E_{\mathrm{ST}}$ & 0.41 &  &    &  0.23 &  0.20 &  0.21 &             &  0.82 &  0.36 \\

\textbf{ACRFLCN}\\
$S_\mathrm{1}$ & 3.03 &  1.00 &    &  2.82 &  1.79 &  3.21 &             &  3.15 &  2.66 \\
$T_\mathrm{1}$ & 2.96 &  1.00 &    0.12 &  2.81 &  1.78 &  3.20 &        0.11 &  3.06 &  2.65 \\
$\Delta E_{\mathrm{ST}}$ & 0.07 &  &  &  0.01 &  0.01 &  0.01 &             &  0.09 &  0.01 \\

\textbf{MAD}\\
$S_\mathrm{1}$ &      &      &          &  0.67 &  1.05 &  0.36 &             &  0.24 &  0.30 \\
$T_\mathrm{1}$ &      &      &         &  0.38 &  0.81 &  0.25 &             &  0.14 &  0.26 \\
$\Delta E_{\mathrm{ST}}$ &      &      &  &  0.29 &  0.24 &  0.23 &             &  0.26 &  0.04 \\

 \hline \hline
\end{tabular*}
\end{threeparttable}
\end{table*}

\clearpage

\bibliographystyle{apsrev4-1}
\bibliography{references}